\RequirePackage[mathlines]{lineno} %newadd: display linenumer

\documentclass[a4paper,11pt]{article}
\usepackage{jheppub} % for details on the use of the package, please
\usepackage[T1]{fontenc} % if needed

\usepackage{multirow}
\usepackage{subfigure}
\usepackage{float}
\usepackage{rotating}
\usepackage{color}

\newcommand{\pip}{\pi^+}
\newcommand{\pim}{\pi^-}
\newcommand{\piz}{\pi^0}

\newcommand{\etac}{\eta_c}

\newcommand{\hc}{h_{c}}
\newcommand{\psip}{\psi(2S)}

\newcommand{\jpsi}{J/\psi}

\newcommand{\EE}{e^+e^-}
\newcommand{\ee}{e^+e^-}
\newcommand{\MM}{\mu^+\mu^-}

\newcommand{\pp}{\pi^+\pi^-}

\newcommand{\pppsip}{\pi^+\pi^- \psip}
\newcommand{\pphc}{\pp h_{c}}

\newcommand{\beq}{\begin{equation}}
\newcommand{\eeq}{\end{equation}}
\newcommand{\bitm}{\begin{itemize}}
\newcommand{\eitm}{\end{itemize}}

\newcommand{\gev}{\mathrm{GeV}}

\newcommand{\dsds}{D^{*+}D^{*-}}
\newcommand{\dsd}{D^{*+}D^{-}}
\newcommand{\dds}{D^{+}D^{*-}}
\newcommand{\dsp}{D^{*+}}
\newcommand{\dsm}{D^{*-}}
\newcommand{\dplus}{D^{+}}
\newcommand{\dm}{D^{-}}
\newcommand{\dz}{D^{0}}
\newcommand{\dzb}{\bar{D}^{0}}

\newcommand{\km}{K^{-}}
\newcommand{\ppjpsi}{\pi^{+}\pi^{-}J/\psi}

\newcommand{\dsz}{D^{*0}}
\newcommand{\dszb}{\bar{D}^{*0}}
\newcommand{\chicz}{\chi_{c0}}

\newcommand{\chicj}{\chi_{cJ}}

\title{\boldmath Cross section measurements of the $\EE\to D^{*+}D^{*-}$ and 
$\EE\to D^{*+}D^{-}$ processes at center-of-mass energies from 4.085 to 4.600~GeV}

\collaborationImg{\includegraphics[angle=90,width=0.15\paperwidth]{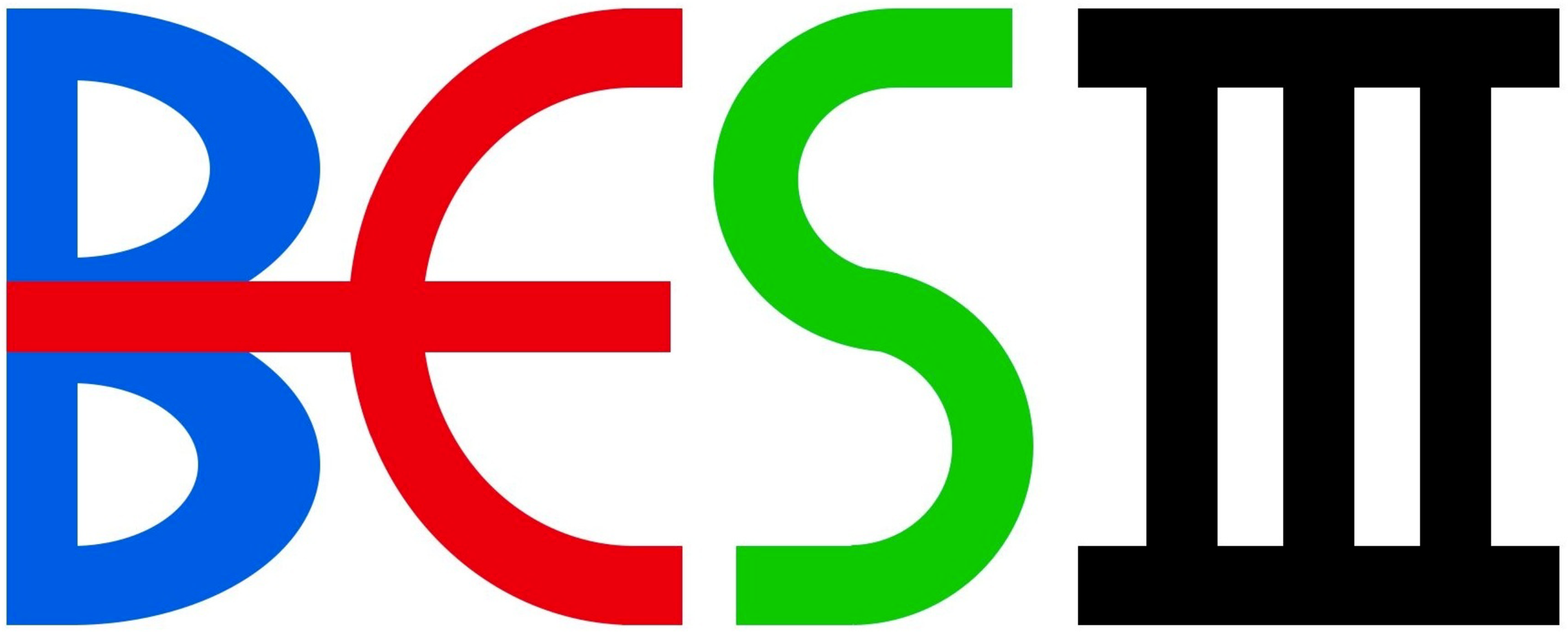}}

\collaboration{The BESIII Collaboration}
 \author{
M.~Ablikim$^{1}$, M.~N.~Achasov$^{10,b}$, P.~Adlarson$^{68}$, M.~Albrecht$^{4}$, R.~Aliberti$^{28}$, A.~Amoroso$^{67A,67C}$, M.~R.~An$^{32}$, Q.~An$^{64,50}$, X.~H.~Bai$^{58}$, Y.~Bai$^{49}$
% ...

\vspace{0.2cm} {\it
$^{1}$ Institute of High Energy Physics, Beijing 100049, People's Republic of China\\
% ...
$^{4}$ Bochum Ruhr-University, D-44780 Bochum, Germany\\
% ...
$^{10}$ G.I. Budker Institute of Nuclear Physics SB RAS (BINP), Novosibirsk 630090, Russia\\
% ...
$^{28}$ Johannes Gutenberg University of Mainz, Johann-Joachim-Becher-Weg 45, D-55099 Mainz, Germany\\
% ...
$^{32}$ Liaoning Normal University, Dalian 116029, People's Republic of China\\
% ...
$^{49}$ Southeast University, Nanjing 211100, People's Republic of China\\
$^{50}$ State Key Laboratory of Particle Detection and Electronics, Beijing 100049, Hefei 230026, People's Republic of China\\
% ...
$^{58}$ University of Jinan, Jinan 250022, People's Republic of China\\
% ...
$^{64}$ University of Science and Technology of China, Hefei 230026, People's Republic of China\\
% ...
$^{67}$ University of Turin and INFN, (A)University of Turin, I-10125, Turin, Italy; (B)University of Eastern Piedmont, I-15121, Alessandria, Italy; (C)INFN, I-10125, Turin, Italy\\
$^{68}$ Uppsala University, Box 516, SE-75120 Uppsala, Sweden\\
% ...
$^{b}$ Also at the Novosibirsk State University, Novosibirsk, 630090, Russia\\
% ...

}
 }

\emailAdd{besiii-publications@ihep.ac.cn}

\abstract{ The Born cross sections of the $\EE\to \dsds$ and $\ee\to\dsd$ processes
are measured  using  $\ee$ collision data collected with the BESIII experiment  at center-of-mass energies from  4.085 to 4.600 GeV, corresponding to an integrated luminosity of $15.7~{\rm fb}^{-1}$. The results are consistent with and more precise than the previous measurements by the Belle, Babar and CLEO collaborations. The measurements are essential for understanding   the nature of vector charmonium and charmonium-like states. }


\arxivnumber{}

\begin{document} 
%\linenumbers%newadd: display linenumer
\maketitle
\flushbottom

\section{Introduction}
\label{sec:intro}

The masses and decay patterns of  conventional vector ($J^{PC}=1^{--}$) charmonium states match well  the predictions from the quark potential model~\cite{barnes}.  They  decay dominantly into open-charm final states when their masses are above the open-charm threshold, such as $\psi(4040)$ and $\psi(4160)$. However, puzzles come when numerous vector charmonium-like states not expected in the quark potential model, such as $Y(4260)$~\cite{intro-BaBar-Y4260, intro-BaBar-Y4260-2012, intro-Belle-Y4260, intro-Belle-Y4260-2} and $Y(4360)$~\cite{intro-BaBar-Y4360, intro-BaBar-Y4360-Y4660-2014, intro-Belle-Y4360-Y4660}, have been observed. 
Later, more precise data indicates that the line shape of $Y(4260)$ is asymmetric and its mass is close to 4220 MeV/$c^2$~\cite{ABLIKIM-2017B, bes3-Y4230-pipihc,bes3-Y4230-omegachic0}.
The measured masses of these charmonium-like states are  above the open-charm threshold. In contrast to conventional states at the same energy region, they mainly decay into hidden-charm final states. %and have  not been observed in any measurements of inclusive hadronic cross sections of $\ee$ annihilation except in Ref.~\cite{intro-Y4220-bes3-open-charm}.
However, the state around 4220 MeV/$c^2$ which was observed in open-charm production process $\ee\to Y(4220)\to\pip D^{0}\bar{D}^{*-}$  in the BESIII experiment~\cite{intro-Y4220-bes3-open-charm}  is consistent with previous observations of the $Y(4220)$ state~\cite{ABLIKIM-2017B,bes3-Y4230-pipihc,bes3-Y4230-omegachic0}. 
A closer examination of these states in open-charm channels may provide further insights on the nature of these states and offer necessary inputs to different theoretical interpretations including  
compact tetraquarks, molecules, hybrids, or hadrocharmonia~\cite{theory-Y-states-chenhuaxing-2016, theory-Y-states-Esposito-2017, theory-Y-states-Richard-2017, theory-Y-states-Ali-2017, theory-Y-states-Stephen-2018,  theory-Y-states-guofenghun-2018, theory-Y-states-Brambilla-2020}.

The previously measured cross sections of $\ee\to\dsds$ and $\ee\to\dsd$ at energy points from 3.875 to 6 GeV~\cite{DstarDstar-belle-2007,DstarDstar-belle-2018, DstarDstar-babar-2009, DstarDstar-cleo-2009} have been used to understand the vector charmonium(-like) states. 
Authors of Ref.~\cite{zhaoqiang2018} found that the cross sections of the $\ee\to\dsds$ process can be described well by the conventional charmonium states $\psi(4040)$, 
$\psi(4160)$, $\psi(4415)$, and a $\bar{D}D_{1}(2420)$ hadronic molecule hypothesis of the $Y(4260)$. Authors of Ref.~\cite{Uglov} used a coupled-channel approach to perform  a simultaneous fit to the data in the open-charm channels including $\ee\to\dsds$ and $\ee\to\dsd$. The fit provides a remarkably good overall description of all the line shapes, with  five vector charmonium resonances, $\psip$, $\psi(3770)$, $\psi(4040)$, $\psi(4160)$, and $\psi(4415)$. Recently, extensive numerical analyses were performed in Ref.~\cite{caoqinfang} by combining all the cross sections of the final states $\jpsi\pp$, $\hc\pp$, $\dz\dsm\pip $, $\psip\pp$, $\omega\chicz$, and $\jpsi\eta$ measured in the BESIII, Belle, and BaBar experiments, together with those of the open-charm final states $\dsds$ and $\dsd$ measured by Belle (with and without $D^{*+}_{s}D^{*-}_{s}$ channel). It is concluded that the $Y(4260)$ may be interpreted as a mixture of $4^{3}S_{1}$ and $3^{3}D_{1}(2^{3}D_{1})$ charmonium states. However, more precise measurements are necessary to verify these conclusions.
         
In this article, we present improved measurements of the Born cross sections of the $\EE\to \dsds$ and $\ee\to\dsd$ processes at 28 center-of-mass (c.m.) energies 
($\sqrt{s}$) from 4.085 to 4.600~GeV.

\section{BESIII detector and data samples}
\label{sec:detector}

The data were collected by the BESIII detector~\cite{Ablikim-2009aa}, and the total integrated luminosity is 15.7~fb$^{-1}$. The c.m.\ energies were measured using 
$\EE\to \MM$ events with an uncertainty of 0.8~MeV~\cite{bes3-energy-measurement} and the integrated luminosities were measured using Bhabha scattering 
events to an uncertainty of 1.0\%~\cite{luminosity-measurement, luminosity-measurement2}. 
The data samples used in this analysis and the corresponding integrated luminosities are listed in Tables~\ref{table:cross-sig} and~\ref{table:cross-bk4}.

 \begin{sidewaystable*}[htbp]
%\vspace{8cm}
\begin{center}
%\linespread{1.5}
%\setlength{\tabcolsep}{7mm}{
\resizebox{22cm}{7cm}{
\begin{tabular}{|c|rr@{}lccr@{}lr@{}lccr@{}lr@{}l|}
%\begin{tabular}{|c|cr@{}lccr@{}l@{}lr@{}lccr@{}l@{}lr@{}l|}
  \hline $\sqrt s$~(GeV) &$ \mathcal{L}_{\rm int}$~(pb$^{-1}$) &\multicolumn{2}{c}{$N^{\rm sig}_{D^{*+}}$} & $\epsilon_{D^{*+}}$ &$(1+\delta)_{D^{*+}}|1-\Pi|^{-2}$ &\multicolumn{2}{c}{$\sigma^{\rm B}_{D^{*+}}$ ~(pb)}  & \multicolumn{2}{c}{$N^{\rm sig}_{D^{*-}}$}& $\epsilon_{D^{*-}}$ &$(1+\delta)_{D^{*-}}|1-  \Pi|^{-2}$ &\multicolumn{2}{c}{$\sigma^{\rm B}_{D^{*-}}$ ~(pb)} &  \multicolumn{2}{c|}{ $\sigma^{\rm B}_{a}$~(pb)}\\\hline
4.0854&52.9&188&$\pm$15&0.055&0.845&2867&$\pm$228$\pm$174&194&$\pm$16&0.054&0.838&3009&$\pm$241$\pm$144&2949&$\pm$259\\
4.1285&393.4&2374&$\pm$53&0.082&0.863&3205&$\pm$72$\pm$168&2379&$\pm$53&0.083&0.874&3121&$\pm$70$\pm$159&3158&$\pm$178\\
4.1574&406.9&3086&$\pm$61&0.100&0.893&3177&$\pm$63$\pm$164&2972&$\pm$60&0.100&0.897&3052&$\pm$61$\pm$154&3106&$\pm$170\\
4.1780&3189.0&25370&$\pm$171&0.120&0.935&2652&$\pm$18$\pm$122&25697&$\pm$171&0.120&0.932&2701&$\pm$18$\pm$137&2662&$\pm$126\\
4.1886&570.0&4358&$\pm$72&0.121&0.946&2496&$\pm$41$\pm$119&4276&$\pm$71&0.122&0.958&2394&$\pm$40$\pm$119&2448&$\pm$126\\
4.1989&526.0&3608&$\pm$65&0.125&1.000&2046&$\pm$37$\pm$98&3651&$\pm$66&0.126&0.987&2081&$\pm$37$\pm$104&2061&$\pm$107\\
4.2092&572.1&3318&$\pm$64&0.128&1.057&1600&$\pm$31$\pm$84&3440&$\pm$65&0.128&1.037&1689&$\pm$32$\pm$91&1638&$\pm$92\\
4.2171&569.2&2653&$\pm$59&0.127&1.208&1136&$\pm$25$\pm$55&2710&$\pm$59&0.124&1.228&1170&$\pm$26$\pm$59&1150&$\pm$62\\
4.2263&1100.9&4150&$\pm$77&0.137&1.365&754&$\pm$14$\pm$36&3890&$\pm$74&0.136&1.463&665&$\pm$13$\pm$32&696&$\pm$36\\
4.2357&530.3&1233&$\pm$48&0.128&2.105&323&$\pm$12$\pm$19&1188&$\pm$45&0.126&1.994&332&$\pm$13$\pm$19&327&$\pm$23\\
4.2438&538.1&1030&$\pm$45&0.123&2.249&258&$\pm$11$\pm$16&954&$\pm$44&0.123&2.313&234&$\pm$11$\pm$14&245&$\pm$19\\
4.2580&828.4&1474&$\pm$52&0.133&2.024&247&$\pm$9$\pm$15&1561&$\pm$57&0.135&2.026&258&$\pm$9$\pm$15&252&$\pm$17\\
4.2668&531.1&990&$\pm$43&0.135&1.755&295&$\pm$13$\pm$16&1032&$\pm$43&0.134&1.703&318&$\pm$13$\pm$16&308&$\pm$20\\
4.2777&175.7&338&$\pm$23&0.135&1.582&336&$\pm$23$\pm$21&367&$\pm$25&0.136&1.532&375&$\pm$25$\pm$19&357&$\pm$30\\
4.2879&491.5&1074&$\pm$40&0.138&1.312&452&$\pm$17$\pm$23&1079&$\pm$40&0.137&1.314&455&$\pm$17$\pm$24&454&$\pm$29\\
4.3079&45.1&118&$\pm$13&0.157&1.216&511&$\pm$55$\pm$35&158&$\pm$14&0.164&1.094&726&$\pm$65$\pm$45&596&$\pm$68\\
4.3121&492.1&1453&$\pm$43&0.156&1.068&661&$\pm$20$\pm$33&1496&$\pm$44&0.155&1.135&645&$\pm$19$\pm$30&651&$\pm$37\\
4.3374&501.1&1998&$\pm$49&0.169&1.034&854&$\pm$21$\pm$43&1980&$\pm$49&0.168&1.013&869&$\pm$21$\pm$43&861&$\pm$48\\
4.3583&543.9&2550&$\pm$55&0.183&0.999&958&$\pm$21$\pm$49&2555&$\pm$55&0.186&1.008&937&$\pm$20$\pm$44&945&$\pm$50\\
4.3774&522.8&2403&$\pm$53&0.181&1.007&946&$\pm$21$\pm$49&2507&$\pm$54&0.179&0.999&1001&$\pm$21$\pm$51&973&$\pm$54\\
4.3874&55.6&323&$\pm$19&0.195&0.947&1180&$\pm$70$\pm$63&259&$\pm$18&0.190&1.058&868&$\pm$59$\pm$43&969&$\pm$76\\
4.3964&505.0&2462&$\pm$53&0.185&1.030&959&$\pm$21$\pm$46&2403&$\pm$53&0.188&0.997&951&$\pm$21$\pm$50&956&$\pm$52\\
4.4156&1090.7&4963&$\pm$76&0.193&1.044&845&$\pm$13$\pm$40&4963&$\pm$76&0.193&1.060&833&$\pm$13$\pm$40&839&$\pm$42\\
4.4362&568.1&2211&$\pm$52&0.189&1.119&690&$\pm$16$\pm$34&2397&$\pm$54&0.192&1.082&759&$\pm$17$\pm$36&721&$\pm$39\\
4.4671&111.1&327&$\pm$20&0.188&1.261&465&$\pm$29$\pm$25&333&$\pm$21&0.188&1.288&463&$\pm$29$\pm$23&464&$\pm$35\\
4.5271&112.1&280&$\pm$19&0.191&1.246&394&$\pm$27$\pm$22&302&$\pm$20&0.198&1.192&428&$\pm$28$\pm$24&410&$\pm$34\\
4.5745&48.9&145&$\pm$13&0.206&1.130&475&$\pm$44$\pm$24&145&$\pm$13&0.201&1.192&462&$\pm$42$\pm$32&469&$\pm$47\\
4.5995&586.9&1479&$\pm$43&0.204&1.238&373&$\pm$11$\pm$18&1473&$\pm$43&0.202&1.219&381&$\pm$11$\pm$18&377&$\pm$21\\
         \hline  
  \end{tabular}
}
\caption{The Born cross section of $\ee\to\dsds$, $\sigma^{\rm B}$,  
together with the integrated luminosity $\mathcal{L}_{\rm int}$, 
the number of signal events $N^{\rm sig}$, the reconstruction efficiency $\epsilon$, 
and the product of the ISR correction factor and vacuum polarization 
factor $(1+\delta)|1-\Pi|^{-2}$. The subscripts $\dsp$, $\dsm$, and $a$ 
denote the results for the reconstructed $\dsp$ candidates, 
 reconstructed $\dsm$ candidates, and the average values, respectively.
The first uncertainties in $\sigma^{\rm B}_{D^{*+}}$ and 
$\sigma^{\rm B}_{D^{*-}}$ are statistical and the second systematic; 
the uncertainties in $\sigma^{\rm B}_{a}$ include the statistical 
and systematic uncertainties calculated using the same 
method as in Ref.~\cite{combine}.}\label{table:cross-sig}
\end{center}
\end{sidewaystable*}

\begin{sidewaystable*}[htbp]
%\vspace{8cm}
\begin{center}%\resizebox{16.5cm}{6cm}{
\resizebox{22cm}{7cm}{
\begin{tabular}{|c|rr@{}lccr@{}lr@{}lccr@{}lr@{}l|}
%\begin{tabular}{|c|cccccccccl|}
     \hline $\sqrt s$~(GeV) &$ \mathcal{L}_{\rm int}$~(pb$^{-1}$)&\multicolumn{2}{c}{$N^{\rm sig}_{D^{*+}}$} & $\epsilon_{D^{*+}}$ &$(1+\delta)_{D^{*+}}|1-\Pi|^{-2}$ &\multicolumn{2}{c}{$\sigma^{\rm B}_{D^{*+}}$~(pb)}  & \multicolumn{2}{c}{$N^{\rm sig}_{D^{*-}}$}& $\epsilon_{D^{*-}}$ &$(1+\delta)_{D^{*-}}|1-  \Pi|^{-2}$ &\multicolumn{2}{c}{$\sigma^{\rm B}_{D^{*-}}$~(pb)}  &\multicolumn{2}{c|}{$\sigma^{\rm B}_{c}$~(pb)}\\\hline
4.0854&52.9&178&$\pm$14&0.135&0.975&947&$\pm$76$\pm$55&162&$\pm$13&0.131&1.027&850&$\pm$70$\pm$48&1785&$\pm$84\\
4.1285&393.4&1028&$\pm$34&0.141&1.110&627&$\pm$21$\pm$31&1068&$\pm$35&0.141&1.055&681&$\pm$22$\pm$33&1303&$\pm$38\\
4.1574&406.9&1070&$\pm$35&0.160&1.065&581&$\pm$19$\pm$32&1137&$\pm$36&0.156&1.057&632&$\pm$20$\pm$31&1217&$\pm$37\\
4.1780&3189.0&8634&$\pm$101&0.173&1.094&536&$\pm$6$\pm$27&8472&$\pm$100&0.173&1.107&518&$\pm$6$\pm$26&1052&$\pm$27\\
4.1886&570.0&1377&$\pm$40&0.171&1.148&460&$\pm$14$\pm$22&1423&$\pm$41&0.171&1.117&489&$\pm$14$\pm$24&947&$\pm$26\\
4.1989&526.0&1292&$\pm$39&0.178&1.124&462&$\pm$14$\pm$24&1278&$\pm$39&0.175&1.125&460&$\pm$14$\pm$22&923&$\pm$26\\
4.2092&572.1&1280&$\pm$39&0.178&1.180&400&$\pm$12$\pm$20&1278&$\pm$39&0.175&1.168&409&$\pm$12$\pm$20&808&$\pm$23\\
4.2171&569.2&1251&$\pm$39&0.180&1.184&389&$\pm$12$\pm$19&1270&$\pm$38&0.179&1.184&393&$\pm$12$\pm$19&781&$\pm$22\\
4.2263&1100.9&2456&$\pm$54&0.194&1.212&356&$\pm$8$\pm$17&2350&$\pm$53&0.192&1.248&331&$\pm$8$\pm$16&684&$\pm$18\\
4.2357&530.3&1217&$\pm$38&0.194&1.142&385&$\pm$12$\pm$19&1136&$\pm$37&0.192&1.177&356&$\pm$11$\pm$17&738&$\pm$21\\
4.2438&538.1&1157&$\pm$37&0.196&1.172&351&$\pm$11$\pm$17&1188&$\pm$37&0.195&1.154&367&$\pm$12$\pm$18&715&$\pm$21\\
4.2580&828.4&2007&$\pm$49&0.208&1.131&385&$\pm$9$\pm$19&1872&$\pm$47&0.207&1.149&356&$\pm$9$\pm$18&738&$\pm$20\\
4.2668&531.1&1301&$\pm$39&0.208&1.108&397&$\pm$12$\pm$19&1286&$\pm$39&0.207&1.100&396&$\pm$12$\pm$20&793&$\pm$22\\
4.2777&175.7&422&$\pm$22&0.205&1.121&390&$\pm$21$\pm$19&468&$\pm$23&0.209&1.033&464&$\pm$23$\pm$24&835&$\pm$29\\
4.2879&491.5&1161&$\pm$37&0.204&1.112&391&$\pm$13$\pm$19&1166&$\pm$37&0.203&1.128&387&$\pm$12$\pm$19&778&$\pm$22\\
4.3079&45.1&109&$\pm$11&0.224&1.131&355&$\pm$37$\pm$19&101&$\pm$11&0.224&1.141&333&$\pm$35$\pm$20&688&$\pm$37\\
4.3121&492.1&1079&$\pm$36&0.211&1.191&327&$\pm$11$\pm$16&1142&$\pm$37&0.212&1.089&376&$\pm$12$\pm$18&694&$\pm$20\\
4.3374&501.1&1182&$\pm$37&0.223&1.091&360&$\pm$11$\pm$17&1150&$\pm$37&0.224&1.131&339&$\pm$11$\pm$17&699&$\pm$20\\
4.3583&543.9&1342&$\pm$40&0.242&1.112&343&$\pm$10$\pm$16&1320&$\pm$39&0.236&1.124&342&$\pm$10$\pm$17&685&$\pm$19\\
4.3774&522.8&1259&$\pm$38&0.233&1.101&351&$\pm$11$\pm$17&1215&$\pm$38&0.231&1.102&342&$\pm$11$\pm$16&692&$\pm$19\\
4.3874&55.6&136&$\pm$12&0.246&1.131&332&$\pm$30$\pm$16&146&$\pm$13&0.248&1.068&371&$\pm$34$\pm$19&695&$\pm$32\\
4.3964&505.0&1191&$\pm$38&0.238&1.107&334&$\pm$11$\pm$16&1277&$\pm$38&0.240&1.064&370&$\pm$11$\pm$19&693&$\pm$20\\
4.4156&1090.7&2688&$\pm$56&0.245&1.094&344&$\pm$7$\pm$16&2672&$\pm$56&0.246&1.113&334&$\pm$7$\pm$16&677&$\pm$17\\
4.4362&568.1&1444&$\pm$41&0.250&1.073&355&$\pm$10$\pm$17&1387&$\pm$40&0.248&1.096&337&$\pm$10$\pm$16&690&$\pm$19\\
4.4671&111.1&282&$\pm$18&0.253&1.083&345&$\pm$22$\pm$16&265&$\pm$17&0.253&1.114&318&$\pm$21$\pm$15&661&$\pm$24\\
4.5271&112.1&260&$\pm$17&0.261&1.144&290&$\pm$19$\pm$14&257&$\pm$17&0.265&1.116&288&$\pm$19$\pm$14&579&$\pm$22\\
4.5745&48.9&78&$\pm$10&0.255&1.304&177&$\pm$22$\pm$11&104&$\pm$11&0.266&1.175&255&$\pm$26$\pm$14&418&$\pm$25\\
4.5995&586.9&1085&$\pm$36&0.268&1.162&224&$\pm$7$\pm$11&1146&$\pm$37&0.269&1.185&229&$\pm$7$\pm$12&451&$\pm$13\\
   \hline 
  \end{tabular}
}
\caption{The Born cross sections of $\ee\to\dsd$ ($\sigma^{\rm B}_{\dsp}$) 
and $\ee\to\dsm\dplus$~($\sigma^{\rm B}_{\dsm}$),  
together with the integrated luminosity $\mathcal{L}_{\rm int}$, 
the number of signal events $N^{\rm sig}$, the reconstruction efficiency $\epsilon$, 
and the product of the ISR correction factor and vacuum polarization 
factor $(1+\delta)|1-\Pi|^{-2}$. The subscripts $\dsp$, $\dsm$, and $c$ 
denote the results for the reconstructed $\dsp$ candidates, reconstructed $\dsm$ 
candidates, and the combined values, respectively.
 The first uncertainties in $\sigma^{\rm B}_{D^{*+}}$ and 
 $\sigma^{\rm B}_{D^{*-}}$ are statistical and the second systematic; 
 the uncertainties in $\sigma^{\rm B}_{c}$ 
 include the statistical and systematic uncertainties calculated 
 using the same method as in Ref.~\cite{combine}.}\label{table:cross-bk4}
\end{center}
\end{sidewaystable*}

The BESIII detector~\cite{Ablikim-2009aa} records symmetric $e^+e^-$ collisions provided by the BEPCII storage ring~\cite{Yu-IPAC2016-TUYA01}, which operates with 
a peak luminosity of $1\times10^{33}$~cm$^{-2}$s$^{-1}$ in the c.m. energy range from 2.0 to 4.95~GeV~\cite{Ablikim:2019hff}.  
 %BESIII has collected large data samples in this energy region~\cite{Ablikim:2019hff}. 
The cylindrical core of the BESIII detector covers 93\% of the full solid angle and consists of a helium-based multilayer drift chamber~(MDC), a plastic scintillator time-of-flight system~(TOF), and a CsI(Tl) electromagnetic calorimeter~(EMC), which are all enclosed in a superconducting solenoidal magnet providing a 1.0~T magnetic field. The solenoid is supported by an octagonal flux-return yoke with resistive plate counter muon identification modules interleaved with steel. 
The charged-particle momentum resolution at $1~{\rm GeV}/c$ is $0.5\%$, and the d$E$/d$x$ resolution is $6\%$ for electrons
from Bhabha scattering. The EMC measures photon energies with a resolution of $2.5\%$ ($5\%$) at $1$~GeV in the barrel (end cap)
region. The time resolution in the TOF barrel region is 68~ps, while that in the end cap region is 110~ps. The end cap TOF
system was upgraded in 2015 using multi-gap resistive plate chamber technology, providing a time resolution of 60~ps~\cite{etof, etof-2, etof-3}.

Simulated data samples produced with a {\sc geant4}-based~\cite{geant4}  Monte Carlo (MC) package, which includes the geometric description of 
the BESIII detector and the detector response, are used to determine detection efficiencies and to estimate background contributions. 
The simulation models, including the beam energy spread, initial state radiation (ISR), and vacuum polarization in the $e^+e^-$ annihilations are treated with 
the generator {\sc kkmc}~\cite{ref-kkmc, ref-kkmc2}.

The signal MC samples of the $\EE\to \dsds$ and $\ee\to\dsd$ processes with  100,000 events 
are generated using phase-space ({\sc PHSP})~\cite{ref-evtgen, ref-evtgen2} and helicity-amplitude ({\sc HELAMP})~\cite{ref-evtgen, ref-evtgen2} models at each c.m.\ energy, respectively. These signal models describe  data well at each energy point.  
The $\dsp$ meson is reconstructed by using the decay chain  $\dsp\to\pip\dz$, $\dz\to\km\pip$, 
while the $\dsm$ or $\dm$ is not reconstructed exclusively  but is inferred from energy-momentum conservation.  
 Inclusion of charge-conjugated ($c.c.$) states is implicit unless otherwise stated.

Generic MC samples  spread over the complete energy range are used to  analyze the possible background contributions. 
The generic MC sample includes the production of open-charm processes, the ISR production of vector charmonium(-like) states, and the continuum processes incorporated in {\sc kkmc}~\cite{ref-kkmc, ref-kkmc2}. 
The known decay modes are modelled with {\sc evtgen}~\cite{ref-evtgen, ref-evtgen2} using branching fractions taken from the
Particle Data Group~\cite{pdg}, and the remaining unknown charmonium decays are modelled with {\sc lundcharm}~\cite{ref:lundcharm, ref:lundcharm2}. 
Final-state radiation~(FSR) from charged final-state particles is incorporated using  {\sc photos}~\cite{photos}.

In addition, exclusive MC samples with 100,000 events each for the processes $\ee\to\dsp\dzb\pim$, $\dsm\dz\pip$, $\dsp\dszb\pim$, $\dsm\dsz\pip$, 
$\dsp\dm\piz$, $\dz\dm\pip$, and $\dsz\dszb$ (listed in Table~\ref{table:The possible backgrounds}) are generated  with {\sc PHSP} model 
at each c.m.\ energy to study possible background contributions. 
Similarly, only $\dsp$ is reconstructed while others
%the $\dsm$, $\dm$, $\dzb$ and $\dszb$ are not reconstructed  exclusively but 
are inferred  from energy-momentum conservation. 
Here $\ee\to\dsp\dzb\pim$ and $\dsm\dz\pip$, and $\ee\to\dsp\dszb\pim$ 
and $\dsz\dsm\pip$ are two pairs of charge-conjugated modes, which have the 
same input line shapes of cross sections. The background contributions at all the  energy points  are found to be the same. 

  \begin{table}[htbp]
\begin{center}
\begin{tabular}{|llll|}
\hline & \multicolumn{2}{c}{Exclusively generated MC samples}& \\\hline
$\EE \to \dsp\bar{D}^0\pim$,      &$\dsp\to \pip D^0$,  & $D^0\to\km\pip$, & $\bar{D}^0\to$ anything             \\ 
$\EE \to\dsm D^0\pip$,      &$D^0\to \km\pip$& $\dsm\to$ anything&\\
$\EE \to \dsp \bar{D}^{*0} \pim$,     &$\dsp\to\pip D^0$        &$D^0\to\km\pip$, & $\bar{D}^{*0}\to$ anything\\
$\EE \to \dsm D^{*0}\pip$,   &$D^{*0}\to\piz\dz(\gamma\dz)$,  &$D^0\to\km\pip$, & $\dsm\to$ anything\\  
$\EE \to \dsp D^- \pi^0$,&$\dsp\to\pip D^0$        &$D^0\to\km\pip$, & $D^-\to$ anything      \\ 
$\EE \to D^- D^0 \pip$,      &$D^0\to\km\pip$,         & $D^-\to$ anything  &\\
$\EE \to D^{*0}\bar{D}^{*0}$, &$D^{*0}\to$ anything,&$\bar{D}^{*0}\to$anything&  \\ 
 \hline
\end{tabular}
\caption{The decay chains of the exclusively generated MC samples for background studies.}
\label{table:The possible backgrounds}
\end{center}
\end{table}

\section{Event selection and background analysis}
\label{sec:eventselection}

Candidate events with at least two pions with positive charge and at least one kaon with negative charge are selected. The charged tracks are required 
to be well reconstructed in the MDC with a polar angle $\theta$ satisfying $|\!\cos\theta|<0.93$, and the distances of the closest approach to the 
interaction point in $x-y$ plane and $z$ direction of $\ee$ c.m. frame have to be less than 1~cm and 10~cm, respectively. 
The particle identification (PID) of kaons and pions is based on the d$E$/d$x$ and time of flight information. Assumption of a given 
particle identification is based on the larger of the two PID-hypothese probabilities P($h$), $h=K,\pi$.  
Kaon candidates are required to satisfy P($K$) $>$ P($\pi$) and P($K$) $>$ 0.001 with momenta larger than $0.3~\gev/c$. 
Pion candidates are required to satisfy P($\pi$) $>$ P($K$) and P($\pi$) $>$ 0.001. The pions with momenta less than 
$0.3~\gev/c$ are named as $\pi_L^+$, whereas the ones with momenta larger than $0.3~\gev/c$ are named as $\pi_H^+$. At least one  $\pi_L^+$ and one $\pi_H^+$ are required in the final states. 

We assume each input charged track originated from a common vertex, and a kinematic fit is performed to the $\km\pip_H$ candidates, which 
constrains the masses of the $\km\pip_H$ candidates to the known mass of the $D^0$ meson~\cite{pdg}, to improve the track momentum resolution 
and to reduce background events. If there are multiple candidates in one event, we choose the $\km\pip_H\pip_L$ combination with the smallest vertex and kinematic fit $\chi^{2}$ and require $\chi^{2}<200$ for further studies.  
To identify signal candidates that involve the $\dz$ meson, we select events with a $\km\pip_H$ invariant mass before the kinematic fit within a window of three standard deviations ($\pm3\sigma$) around the $\dz$ known mass, $1845.4< M(\km\pip_H)< 1885.2$~MeV/$c^2$, referred to as the $\dz$ mass window. Here $\sigma$ is the measured mass resolution of $\dz$ meson. For the sake of simplicity, we use $\pip$ to replace  $\pi_L^+$ in the following text.

\begin{figure*}[htbp]
\begin{center}
\subfigure{
\includegraphics[width=0.24\paperwidth]
{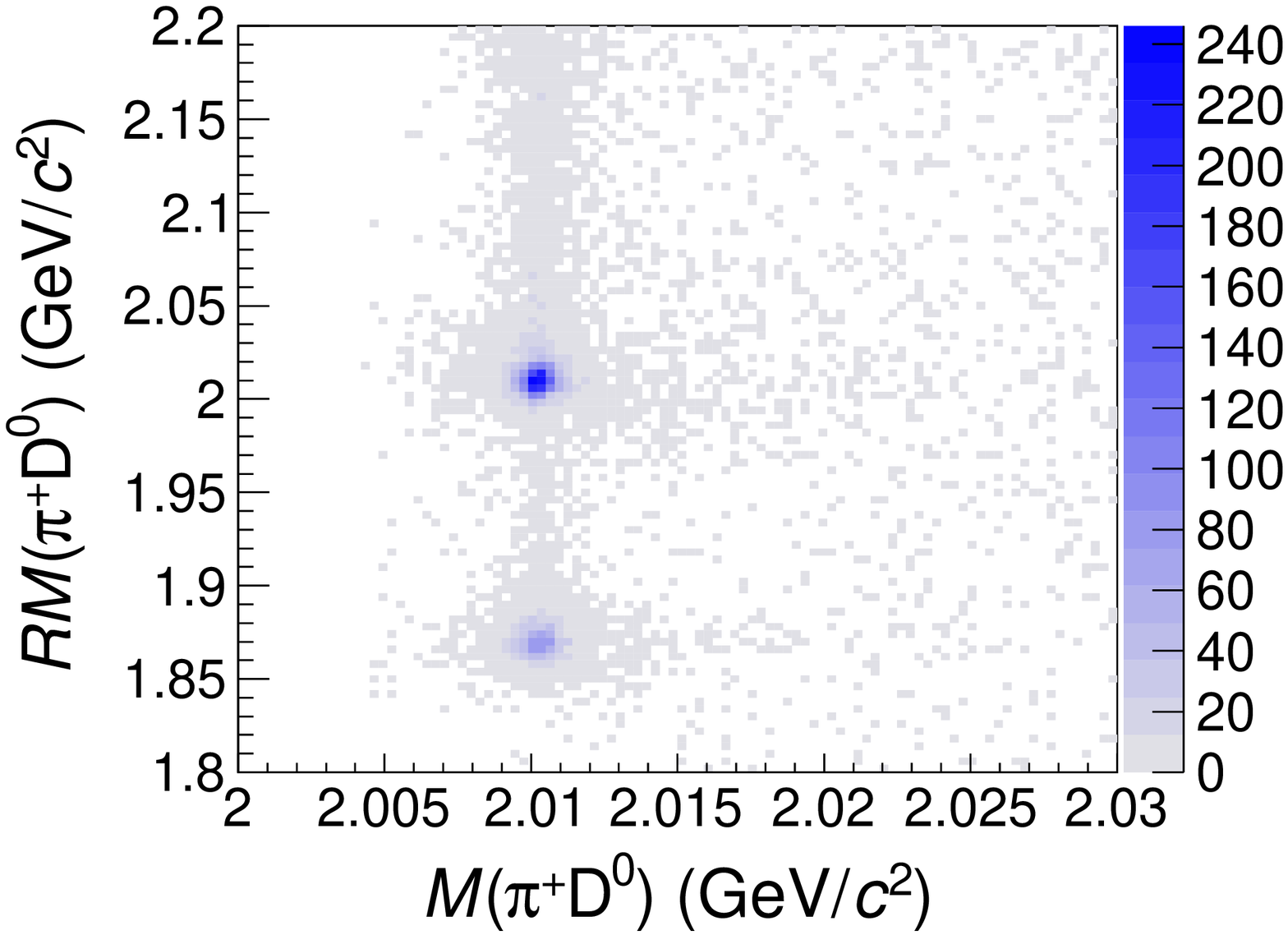}
%\put(-115,30){\textbf{(a)}}
\put(-112,89){\textbf{(a)}}
\put(-112,22){\textbf{Data}}
}
\subfigure{
\includegraphics[width=0.24\paperwidth]
{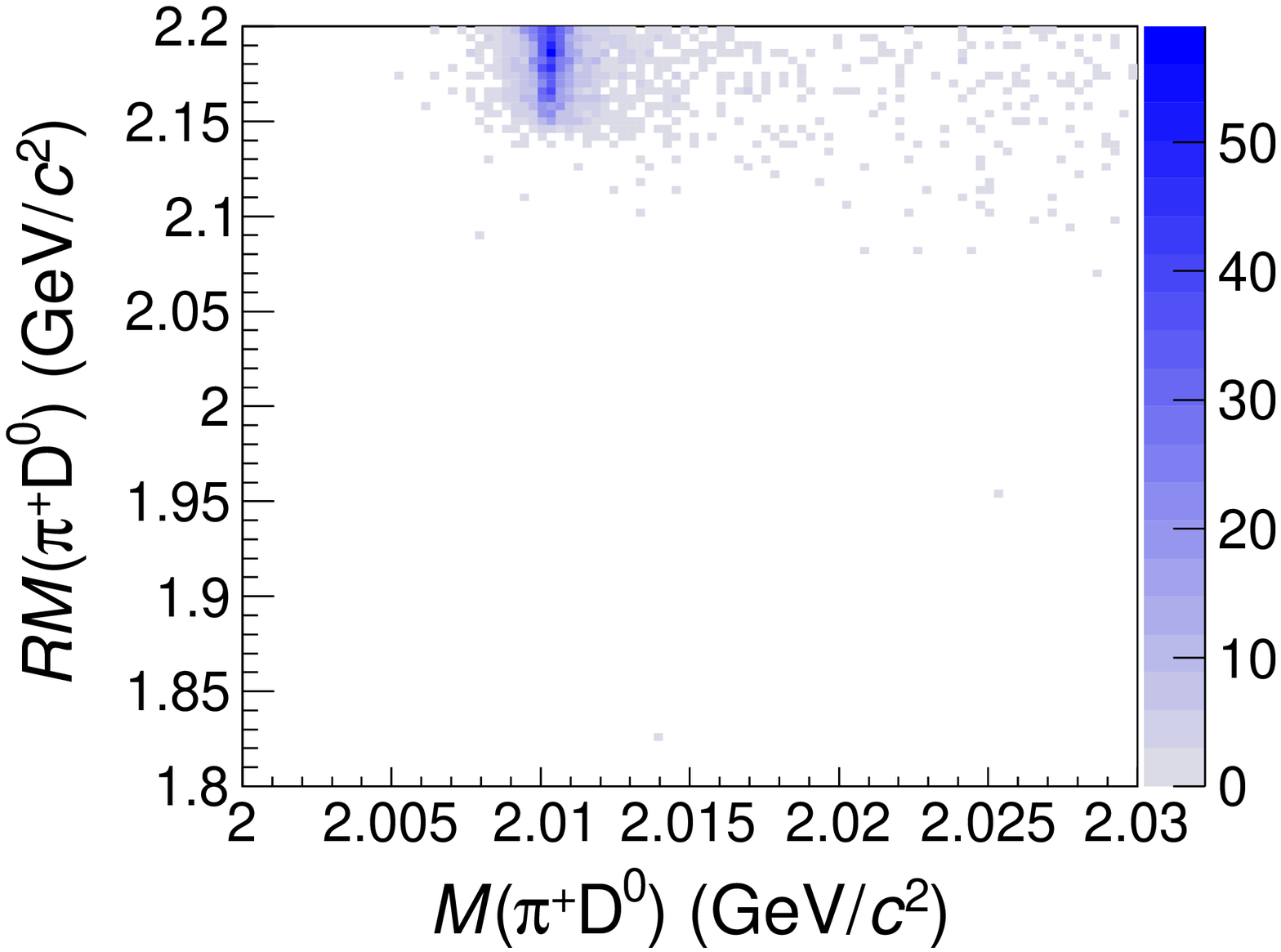}
\put(-112,89){\textbf{(f)}}
\put(-112,35){\textbf{b3}}
\put(-112,22){\textbf{bkg2}}
}\\
\subfigure{
\includegraphics[width=0.24\paperwidth]
{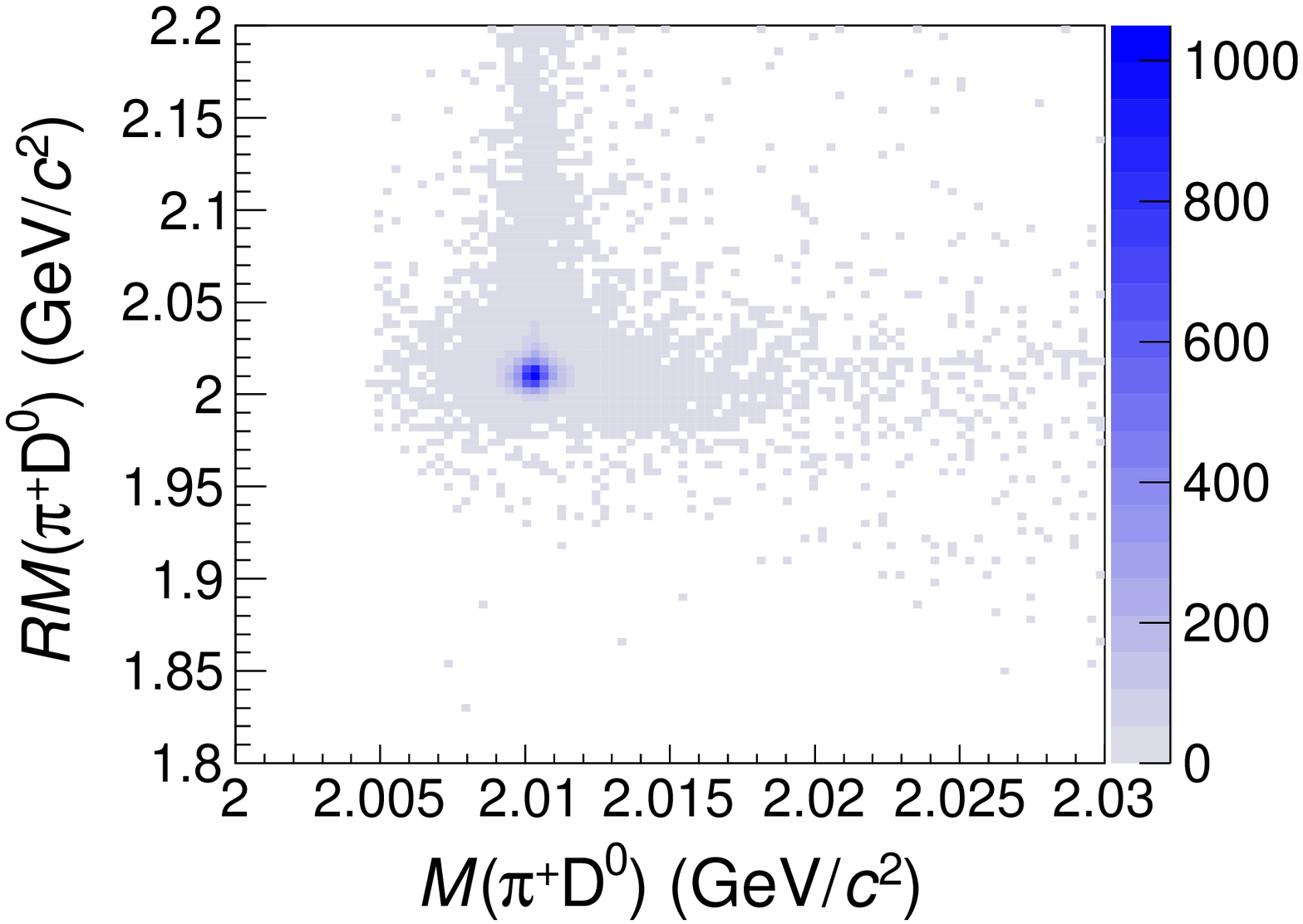}
\put(-112,89){\textbf{(b)}}
\put(-112,35){\textbf{s1}}
\put(-112,22){\textbf{sig1}}
}
\subfigure{
\includegraphics[width=0.24\paperwidth]
{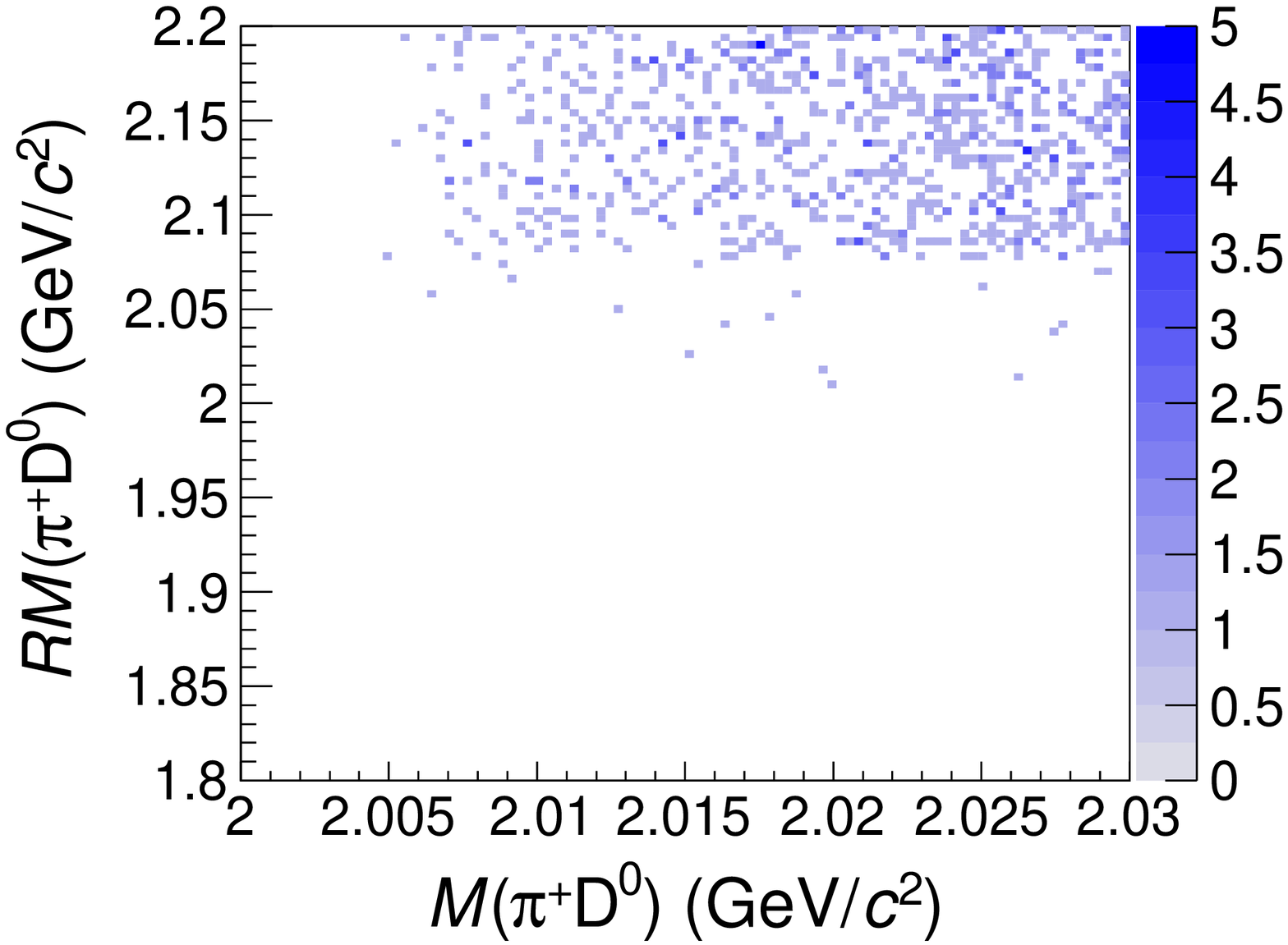}
\put(-112,89){\textbf{(g)}}
\put(-112,35){\textbf{b4}}
\put(-112,22){\textbf{bkg2}}
}\\
\subfigure{
\includegraphics[width=0.24\paperwidth]
{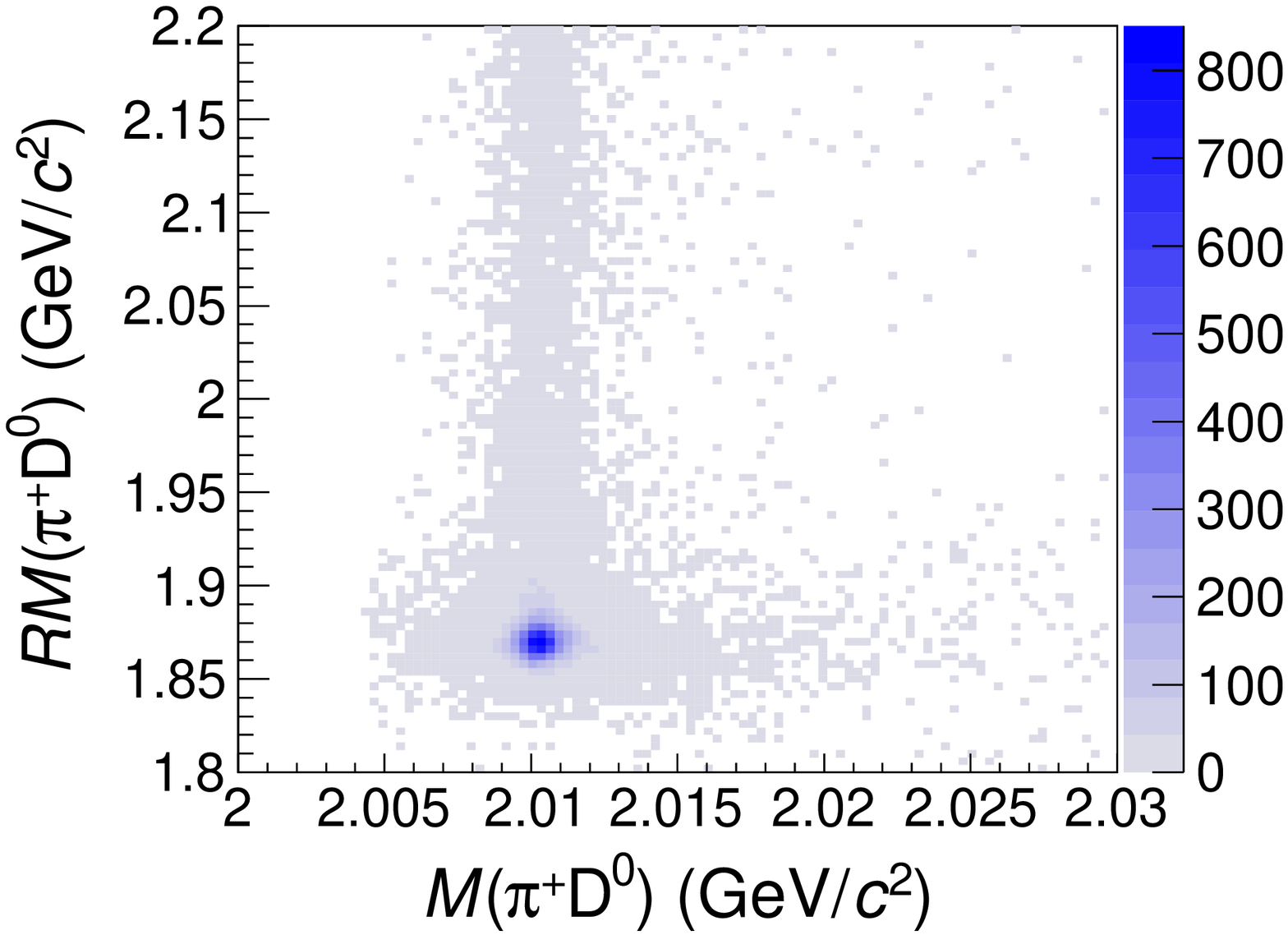}
\put(-112,89){\textbf{(c)}}
\put(-112,22){\textbf{sig2}}
\put(-112,35){\textbf{s2}}
}
\subfigure{
\includegraphics[width=0.24\paperwidth]
{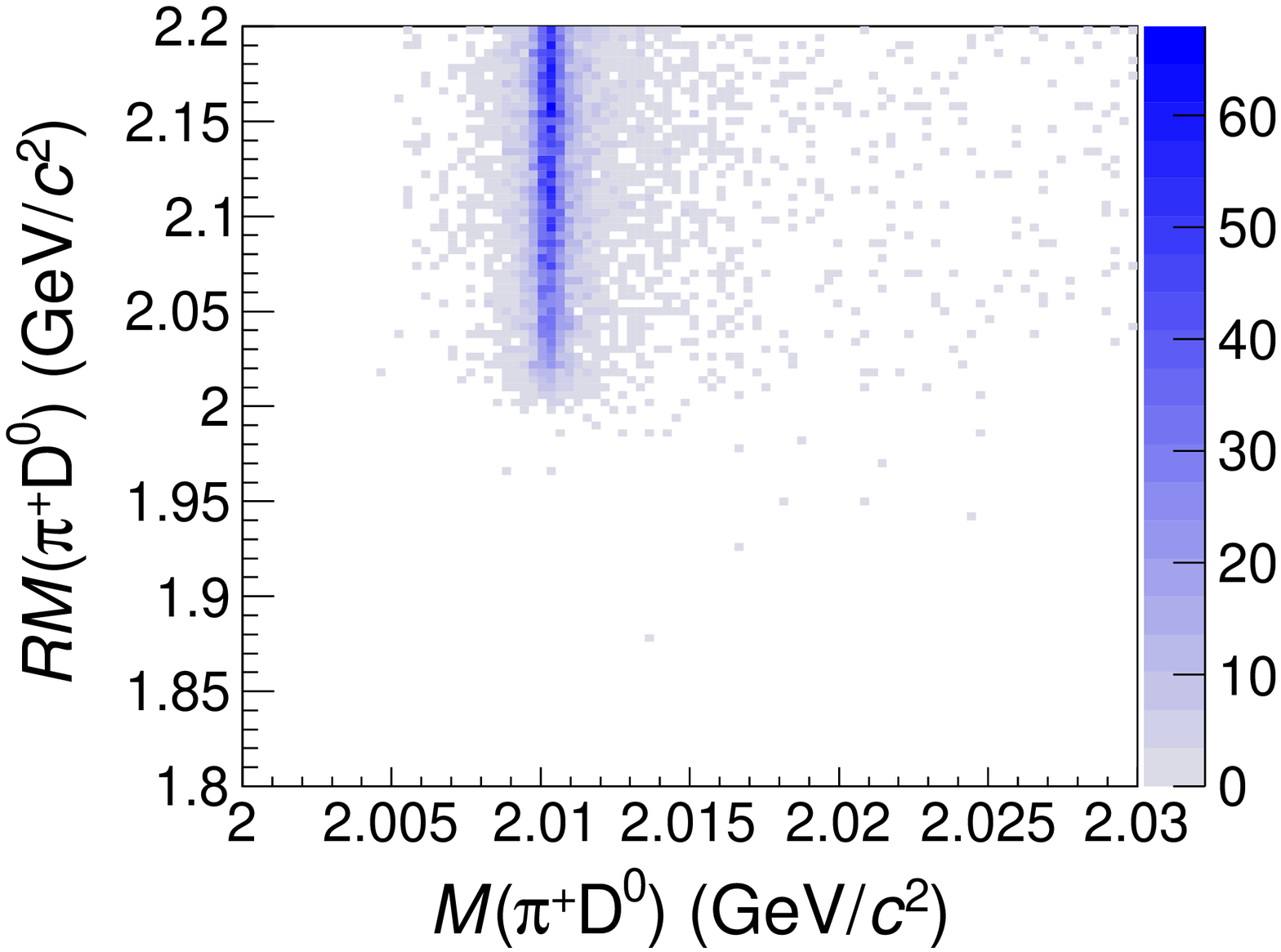}
\put(-112,89){\textbf{(h)}}
\put(-112,35){\textbf{b5}}
\put(-112,22){\textbf{bkg3}}
}\\
\subfigure{
\includegraphics[width=0.24\paperwidth]
{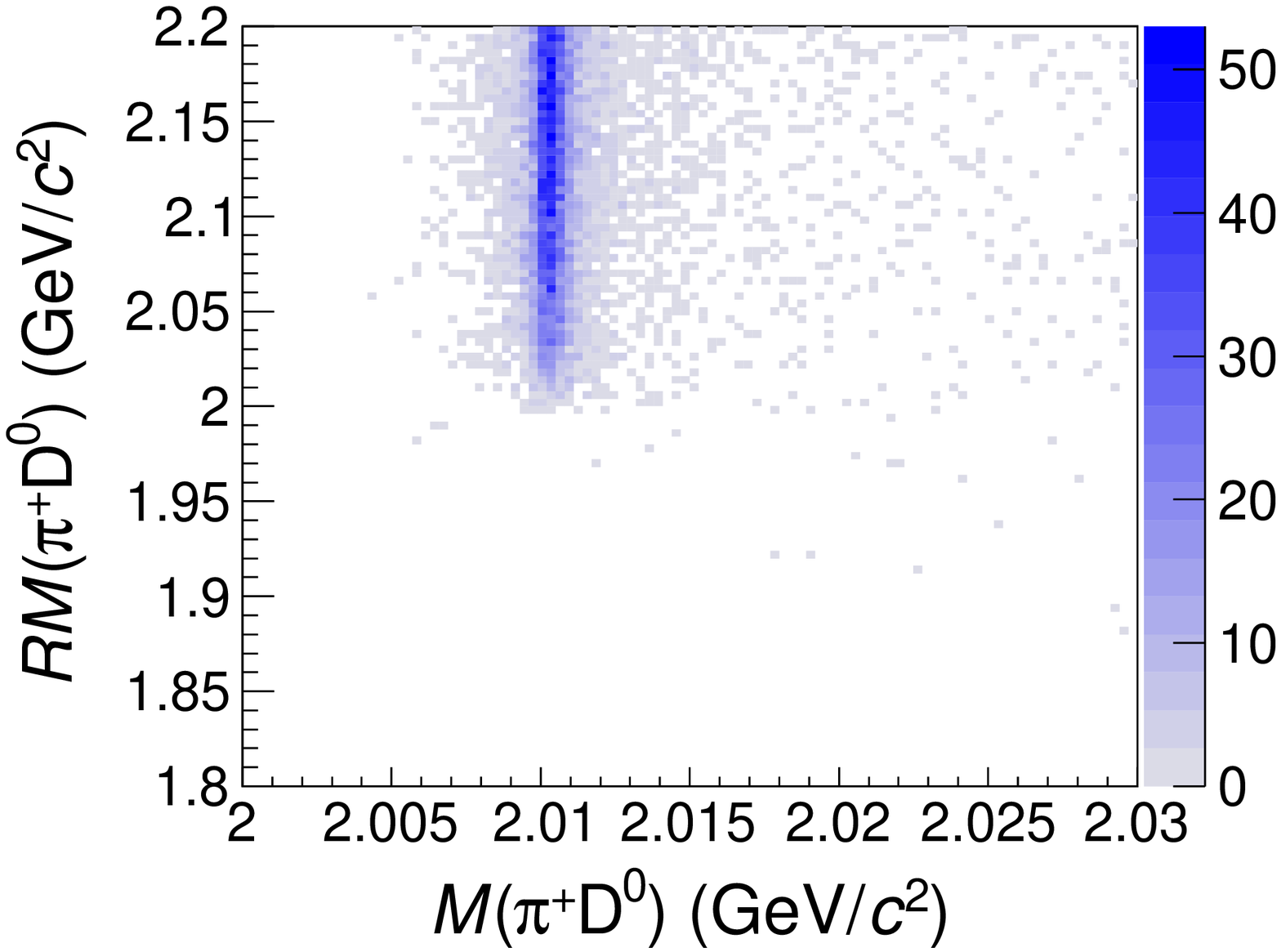}
\put(-112,89){\textbf{(d)}}
\put(-112,35){\textbf{b1}}
\put(-112,22){\textbf{bkg1}}
}
\subfigure{
\includegraphics[width=0.24\paperwidth]
{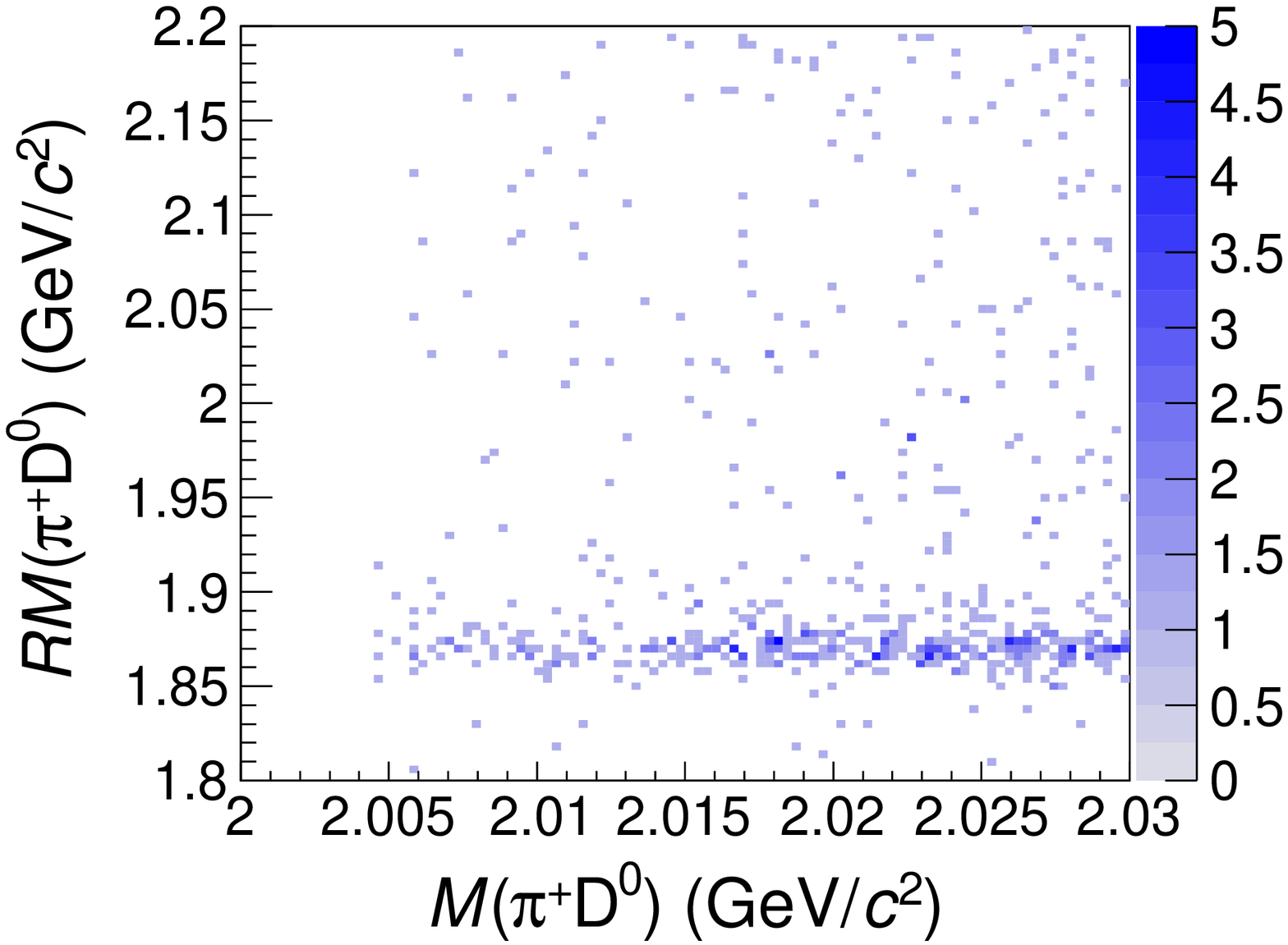}
\put(-112,89){\textbf{(i)}}
\put(-112,35){\textbf{b6}}
\put(-112,22){\textbf{bkg4}}
}\\
\subfigure{
\includegraphics[width=0.24\paperwidth]
{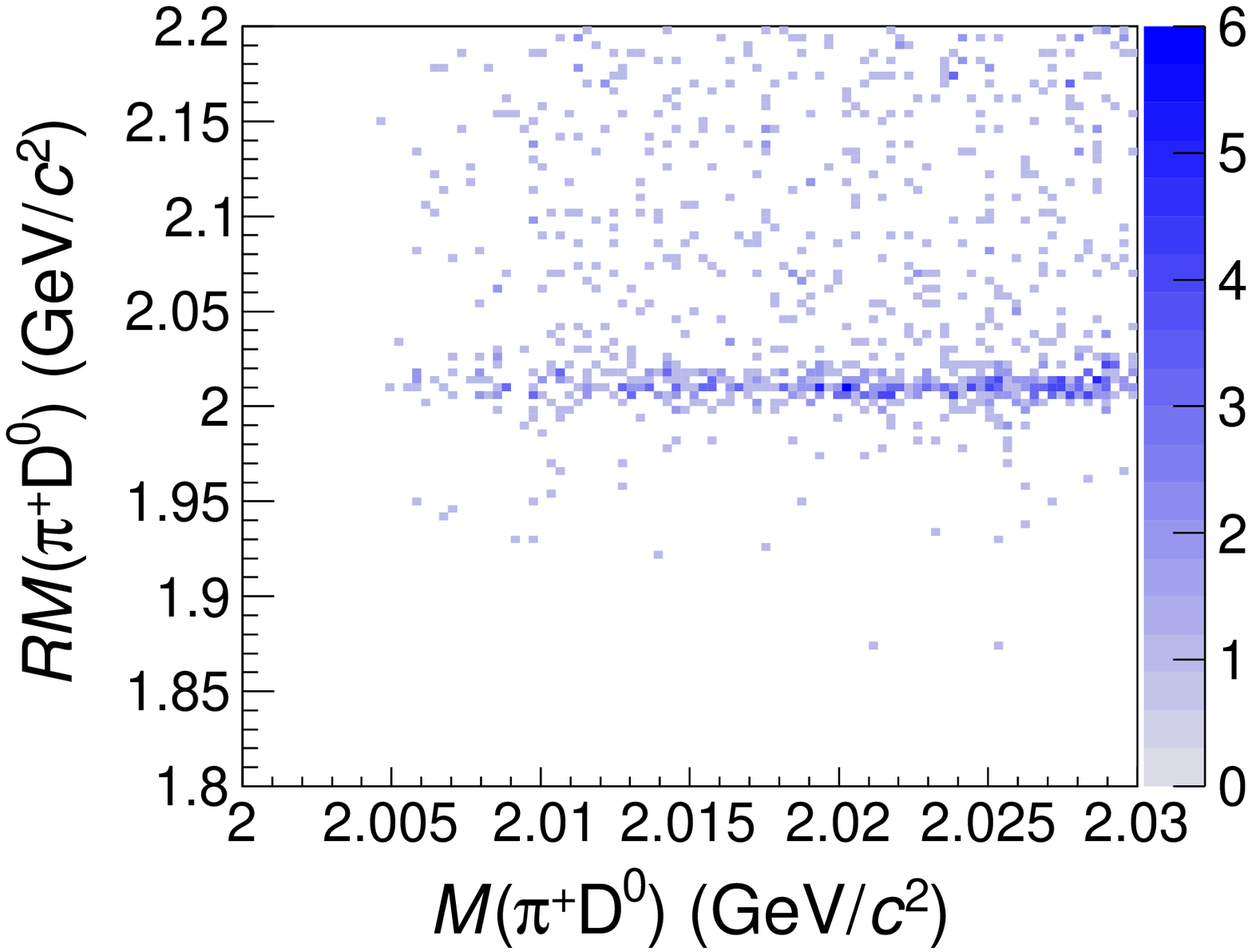}
\put(-112,89){\textbf{(e)}}
\put(-112,35){\textbf{b2}}
\put(-112,22){\textbf{bkg1}}
}
\subfigure{
\includegraphics[width=0.24\paperwidth]
{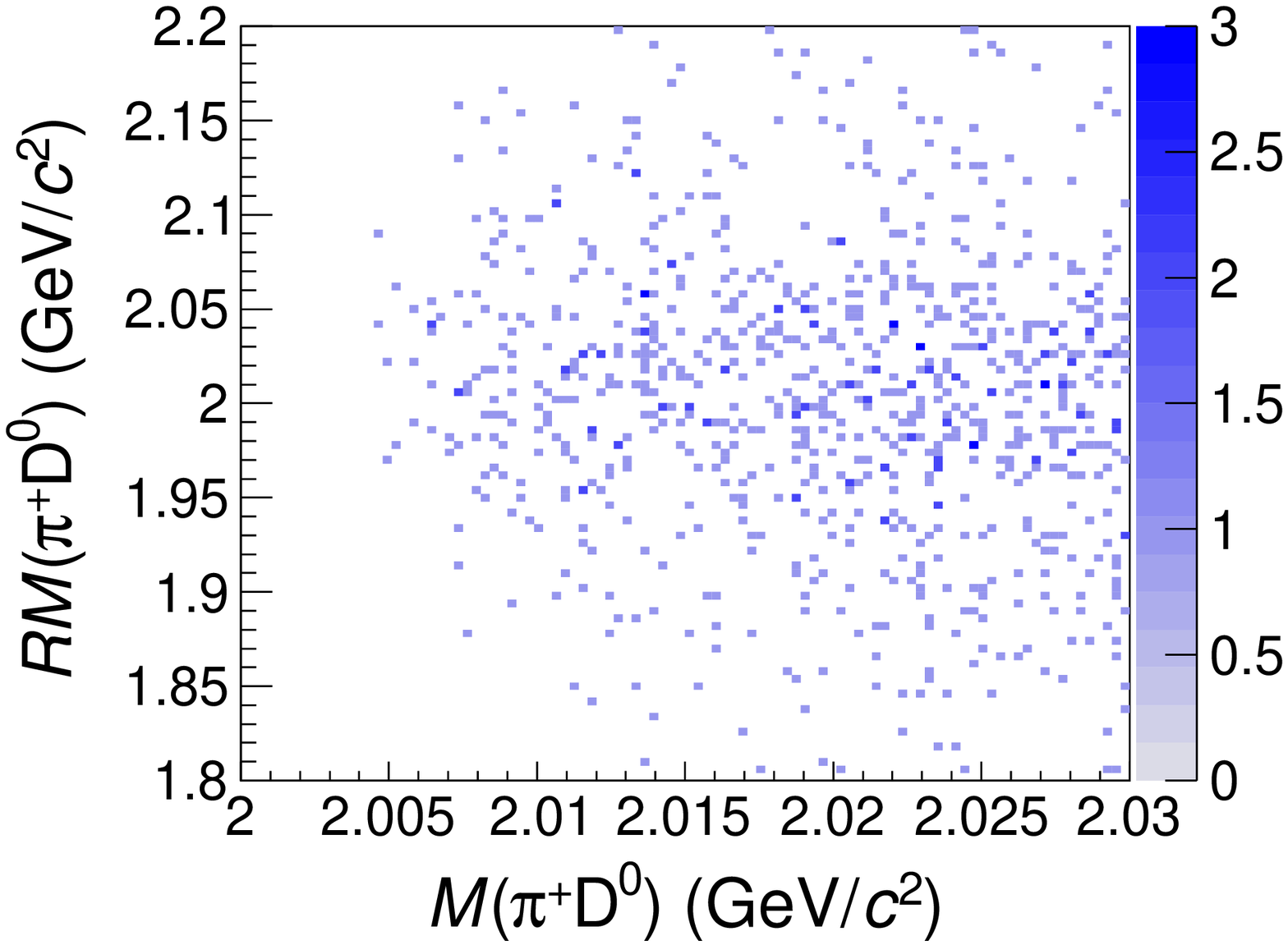}
\put(-112,89){\textbf{(j)}}
\put(-112,35){\textbf{b7}}
\put(-112,22){\textbf{bkg5}}
}\\
\end{center}
\caption{Distributions of  the recoil mass $RM(\pip D^0)$ versus the invariant mass $M(\pip D^0)$  for (a) data, and the MC-simulated processes (b) $\ee\to\dsds$, (c) $\dsd$, (d) $\dsp\bar{D}^0\pim$, (e) $\dsm D^0\pip$, (f) $\dsp \bar{D}^{*0} \pim$, (g) $\dsm D^{*0}\pip$, (h) $\dsp D^- \pi^0$, (i) $D^- D^0 \pip$, and (j) $D^{*0}\bar{D}^{*0}$ 
at 4.416~GeV c.m.\ energy, where plots (b) and (c) are the signal processes, and plots (d-j) are the background processes in this analysis. Labels of s1, sig1, b1, bkg1 etc. are used in eq.(\ref{equ:2dfit}) and described in text.}
\label{fig:scatter inmc hadrons single}
\end{figure*}

After imposing all the requirements mentioned above, we use the two-dimensional~(2D) distributions of the $\pip\dz$ invariant mass $M(\pip\dz)$  and 
recoil mass $RM(\pip\dz)$ after the kinematic fit,  to study the signal and background contributions. 
Figure~\ref{fig:scatter inmc hadrons single} shows the distributions of $RM(\pip\dz)$ versus $M(\pip\dz)$ for data,  signal MC samples of
$\ee\to\dsds$ and $\dsd$, and background MC samples of $\ee\to\dsp\dzb\pim$,
$\dsm\dz\pip$, $\dsp\dszb\pim$, $\dsz\dsm\pip$, $\dsp\dm\piz$, 
$\dz\dm\pip$, and $\ee\to\dsz\dszb$ at 4.416~GeV, respectively.

\begin{figure*}[htbp]
\begin{center}
\subfigure{
\includegraphics[width=0.23\paperwidth]
{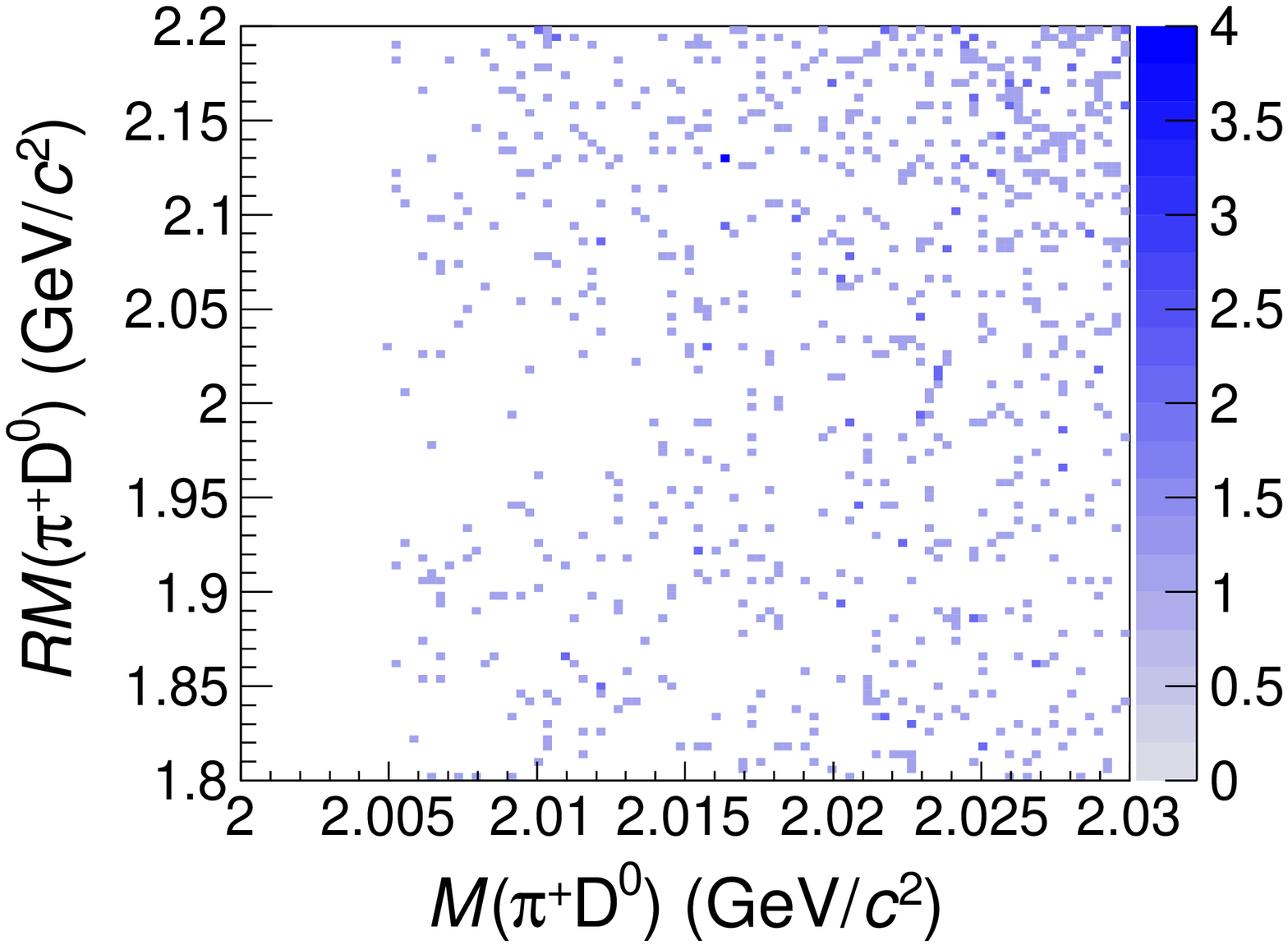}
\put(-105,80){\textbf{bkg6}}
}
\subfigure{
\includegraphics[width=0.46\paperwidth]
 {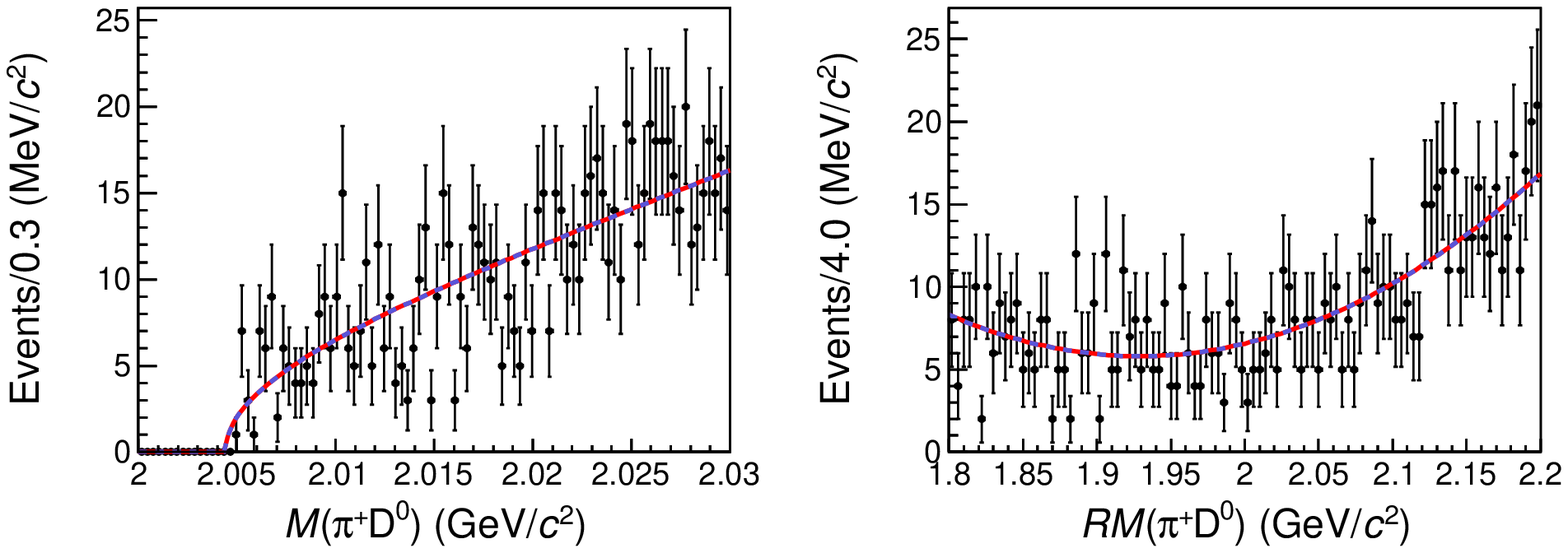}
  \put(-241,80){\textbf{b8}}
 \put(-104,80){\textbf{b9}}
 }
\end{center}
\caption{Distribution of the recoil mass $RM(\pip D^0)$   versus  the invariant mass $M(\pip D^0)$ and the 2D fit result for the remaining background contributions of the generic MC samples after removing signal events and peaking background contributions mentioned in the text at 4.416~GeV c.m.\ energy. An Argus function and a second-order Chebyshev polynomial (red lines) are used to fit the $M(\pip D^0)$ and $RM(\pip D^0)$ distributions, respectively. Labels of bkg6, b8, and b9 are used in eq.(\ref{equ:2dfit}) and described in text.}
\label{fig:scatter inmc hadrons smooth}
\end{figure*}

%The distributions of $M(\pip D^0)$ and $RM(\pip D^0)$ are flat for the generic MC sample and therefore are described with linear functions, when the above signal 
%and background processes are removed using a generic event type analysis tool, TopoAna~\cite{topo}. Distribution and the 2D fit result 
%for the $RM(\pip D^0)$ versus $M(\pip D^0)$ distributions from 
The remaining background contributions are shown in
Figure~\ref{fig:scatter inmc hadrons smooth} together with the fit results where an Argus function~\cite{argus} and a second-order Chebyshev polynomial are used to fit the $M(\pip D^0)$ and 
$RM(\pip D^0)$ distributions, respectively.

\section{Signal yield determination}
\label{sec:background}

Two-dimensional unbinned fits to the recoil mass $RM(\pip\dz)$ versus the invariant mass $M(\pip\dz)$ distributions are performed to determine the signal yields of the $\ee\to\dsds$ and $\dsd$ processes.  A 2D probability density function (PDF) $f(M,RM)$ is used to describe the data, where 
\begin{equation}\label{equ:2dfit}
\begin{split}
 f(M,RM)=& N^{\rm sig1}s_1(M,RM)+N^{\rm sig2}s_2(M,RM)\\
 &+N^{\rm bkg1}[\frac{R_1}{1+R_1}b_1(M,RM)+\frac{1}{1+R_1}b_2(M,RM)] \\
 &+N^{\rm bkg2}[\frac{R_2}{1+R_2}b_3(M,RM)+\frac{1}{1+R_2}b_4(M,RM)] \\
 &+N^{\rm bkg3}b_5(M,RM)+N^{\rm bkg4}b_6(M,RM) \\
 &+N^{\rm bkg5}b_7(M,RM)+N^{\rm bkg6}b_8(M)b_9(RM).
 \end{split}
\end{equation}
%and the peaking contributions due to $\dsp$, $\dsm$, or $\dm$, shown in Fig.~\ref{fig:scatter inmc hadrons single}, are explicitly considered.
Here, $s_1(M,RM)$ and $s_2(M,RM)$ are the signal PDFs for the $\ee\to\dsds$ and $\dsd$ processes, respectively, and are modeled using the signal MC shapes convoluted with corresponding Gaussian functions.  
The parameters of the Gaussian functions reflect the differences in the mass resolution  between MC simulation and 
 data, and are obtained from one-dimensional~(1D) fits to the $M(\pip\dz)$  and $RM(\pip\dz)$ distributions. 
 The $b_1(M,RM)$, $b_2(M,RM)$, $b_3(M,RM)$, $b_4(M,RM)$, $b_5(M,RM)$, $b_6(M,RM)$, and $b_7(M,RM)$ are the PDFs of the background processes
$\ee\to\dsp\bar{D}^0\pim$, $\dsm D^0\pip$, $\dsp \bar{D}^{*0} \pim$, $\dsm D^{*0}\pip$, $\dsp D^- \pi^0$, $D^- D^0 \pip$, and $D^{*0}\bar{D}^{*0}$ and are shown in Figure~\ref{fig:scatter inmc hadrons single}(d-j).  
 The $b_8(M)$ and $b_9(RM)$ are the Argus function and second-order Chebyshev polynomial mentioned before. 
 The PDFs for different contributions in eq.(\ref{equ:2dfit}) vary for different c.m.\ energies.
 
 The numbers of the signal events of $\ee\to\dsds$ and $\dsd$, the background events of $\ee\to\dsp D^-\pi^0$, $D^- D^0 \pip$, and $D^{*0}\bar{D}^{*0}$, and the flat background events are represented by $N^{\rm sig1}$, $N^{\rm sig2}$, $N^{\rm bkg3}$, $N^{\rm bkg4}$, $N^{\rm bkg5}$ and $N^{\rm bkg6}$, respectively;   
$N^{\rm bkg1}$ is the total number of $\ee\to\dsp\bar{D}^0\pim$ and $\dsm D^0\pip$, and $N^{\rm bkg2}$  
the total number of  $\ee\to\dsp \bar{D}^{*0} \pim$ and $\dsm D^{*0}\pip$.
  The ratios between the charge-conjugated modes of $\ee\to\dsp\dzb\pim$ and $\dsm\dz\pip$, and $\ee\to\dsp\dszb\pim$ and $\dsz\dsm\pip$ are denoted by $R_1$ and $R_2$, according to $R_1=\mathcal{B}_{\rm bkg1}\epsilon_{\rm bkg1}/\mathcal{B}_{\rm bkg1}^{c.c.}\epsilon_{\rm bkg1}^{c.c.}$ and  
$R_2=\mathcal{B}_{\rm bkg2}\epsilon_{\rm bkg2}/\mathcal{B}_{\rm bkg2}^{c.c.}\epsilon_{\rm bkg2}^{c.c.}$, 
%where $\mathcal{B}_{\rm bkg1}$, $\mathcal{B}_{\rm bkg1}^{c.c.}$, $\mathcal{B}_{\rm bkg2}$, and $\mathcal{B}_{\rm bkg2}^{c.c.}$ are the branching fractions of the processes $\ee\to\dsp\dzb\pim$, $\dsm\dz\pip$, $\dsp\dszb\pim$, and $\dsz\dsm\pip$, respectively, where the branching fractions of the charmed meson decays are included.
where $\mathcal{B}_{\rm bkg1}$, $\mathcal{B}_{\rm bkg1}^{c.c.}$, $\mathcal{B}_{\rm bkg2}$, and $\mathcal{B}_{\rm bkg2}^{c.c.}$ are the products of branching fractions from 
the intermediate states for the processes $\ee\to\dsp\dzb\pim$, $\dsm\dz\pip$, $\dsp\dszb\pim$, and $\dsz\dsm\pip$, respectively.  
 The corresponding  reconstruction efficiencies are $\epsilon_{\rm bkg1}$, $\epsilon_{\rm bkg1}^{c.c.}$, $\epsilon_{\rm bkg2}$, and $\epsilon_{\rm bkg2}^{c.c.}$.
All the numbers of events in the 2D fit are left free. 
The cross feeds from the charge-conjugated modes of signal channels are below 0.02\% and  therefore are neglected.

\begin{figure*}[htbp]
\begin{center}
\includegraphics[width=0.5\paperwidth]
{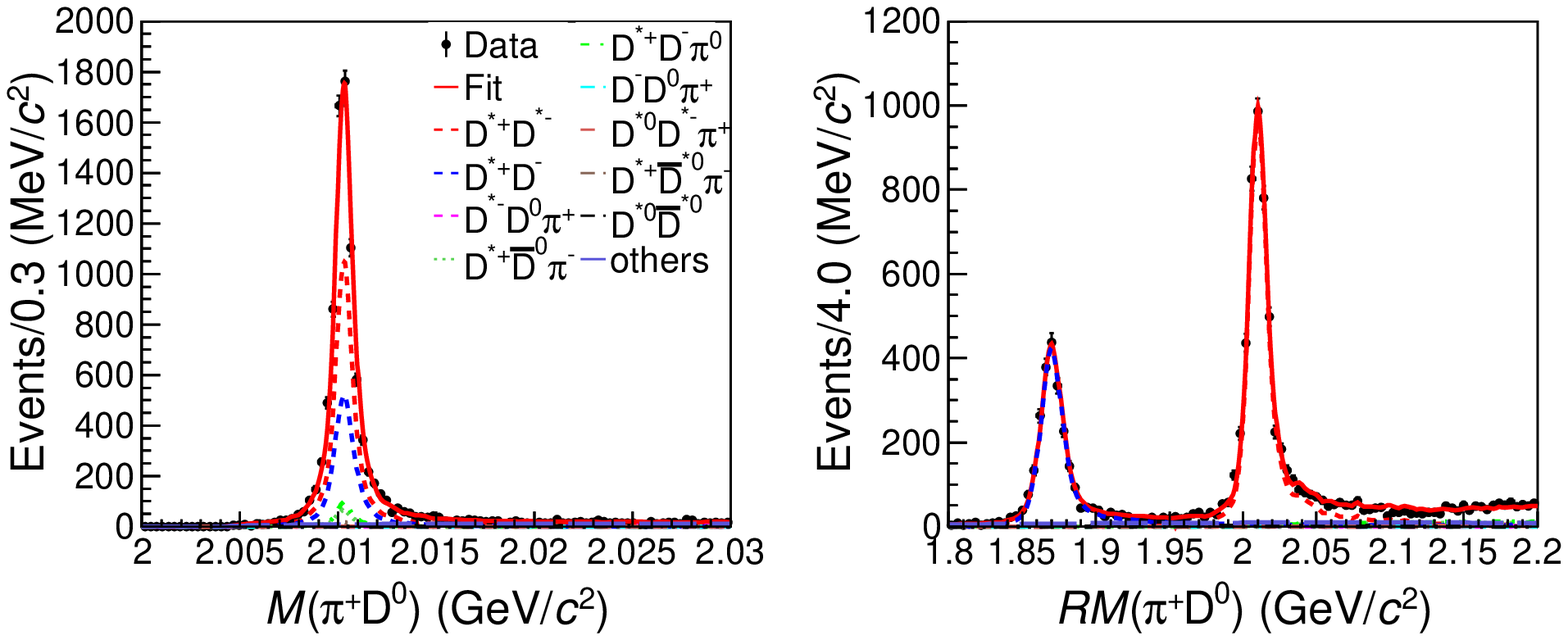}
\includegraphics[width=0.5\paperwidth]
{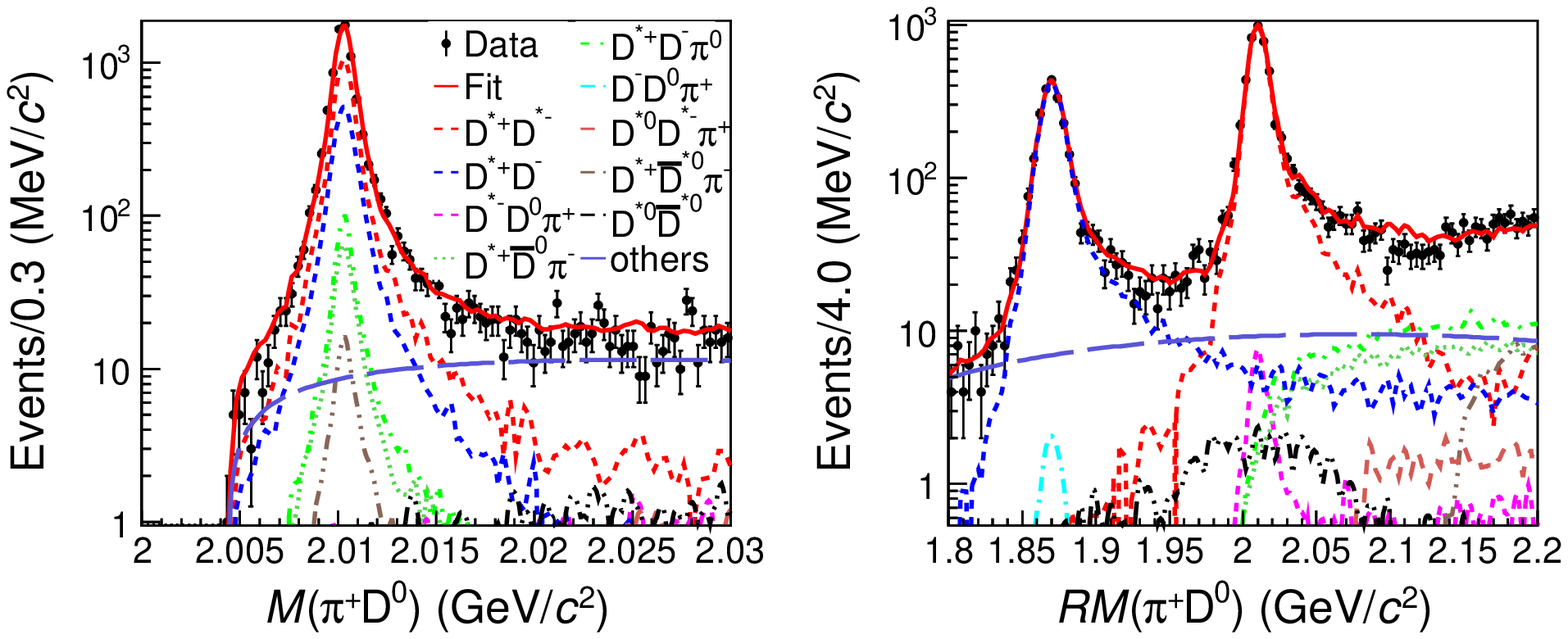}
\end{center}
\caption{The  1D projections of the 2D fit (top) to the $RM(\pip\dz)$ versus $M(\pip\dz)$ distributions for the reconstructed $D^{*+}$ candidates at 4.416~GeV and the corresponding log-scale plots (bottom). Dots with error bars are data. The red solid lines represent the fit result; the red and blue dashed lines denote the signal processes of $\ee\to\dsds$ and $\ee\to\dsd$, 
respectively;   the other dashed lines in different colors represent different background events.    }
\label{fig:2dfit}
\end{figure*}

The 1D projections of the  fit results to the $RM(\pip\dz)$ versus $M(\pip\dz)$  distributions at 4.416~GeV and the corresponding log-scale plots are shown in Figure~\ref{fig:2dfit}.
 The distributions at other energy points are fitted using  the same method to determine the signal yields, which
 are summarized in Tables~\ref{table:cross-sig} and~\ref{table:cross-bk4}.

\section{Cross section measurements}
\label{sec:cross}

The Born cross sections of the reaction channels $\ee\to\dsds$ and $\ee\to\dsd$ are calculated with 
\begin{equation}\label{equ:cross section calculated}
 \sigma^{\rm B}=\frac{N^{\rm sig}}
 {\mathcal{L_{\rm int}}(1+\delta)|1-\Pi|^{-2}\mathcal{B}_{1}\mathcal{B}_{2}\epsilon},
\end{equation}
where $N^{\rm sig}$ is the signal yield;
$\mathcal{L_{\rm int}}$ is the integrated luminosity;  
$|1-\Pi|^{-2}$ is the vacuum polarization factor~\cite{ vacuum-polarization-factor}; 
$\mathcal{B}_1$ and $\mathcal{B}_2$ are the branching fractions of $\dsp\to\pip\dz$ and $\dz\to\km\pip$~\cite{pdg}, respectively; 
$\epsilon$ is the reconstruction efficiency for the $\ee\to\dsds$ or $\ee\to\dsd$ mode. 
 The ISR correction factor  $(1+\delta)$ is obtained iteratively from the quantum electrodynamics calculation~\cite{ref-kkmc, ref-kkmc2, isr-calculate2} by
using the {\sc kkmc} generator. 
The  line shapes of Belle results~\cite{DstarDstar-belle-2018} are used as a starting point to  obtain the reconstruction efficiency and the ISR correction factor for each c.m.\ energy.
 Then the line shapes of the measured cross sections of $\ee\to\dsds$ and $\ee\to\dsd$ are taken as inputs  to re-calculate the efficiency and the ISR correction factor. 
This procedure is repeated until the difference of $(1+\delta)\epsilon$ between two subsequent iterations is less than 3\%.
The Born cross sections of $\ee\to\dsds$ and $\ee\to\dsd$, and the numbers used in the calculation are listed in Tables~\ref{table:cross-sig} and~\ref{table:cross-bk4}.

\begin{figure*}[htbp]
\begin{center}
\subfigure{
\label{fig:born cross section a}
\includegraphics[width=0.3\paperwidth]
 {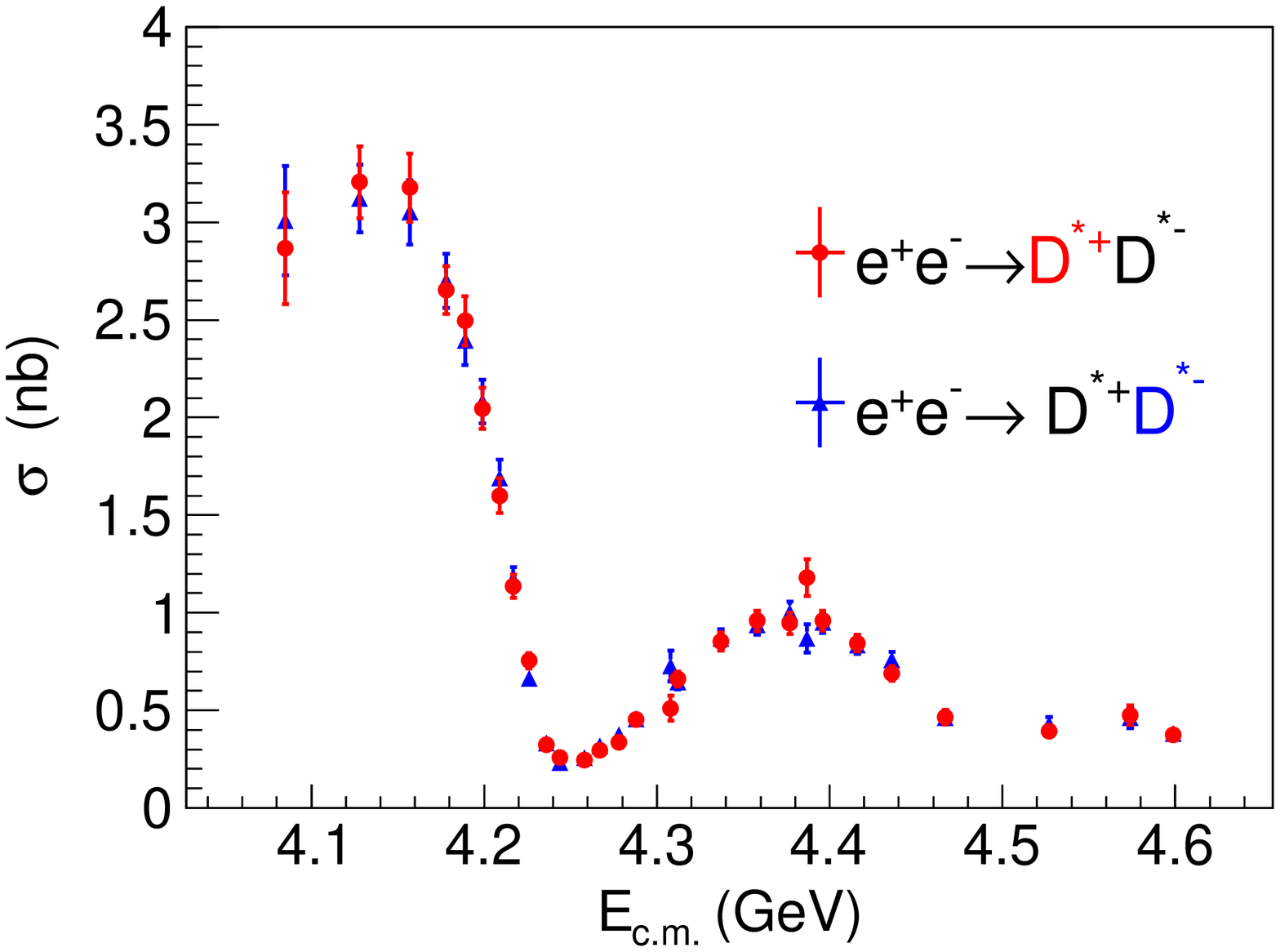}
 \put(-50,110){\textbf{(a)}}
 }
 \subfigure{
\label{fig:born cross section a}
\includegraphics[width=0.3\paperwidth]
 {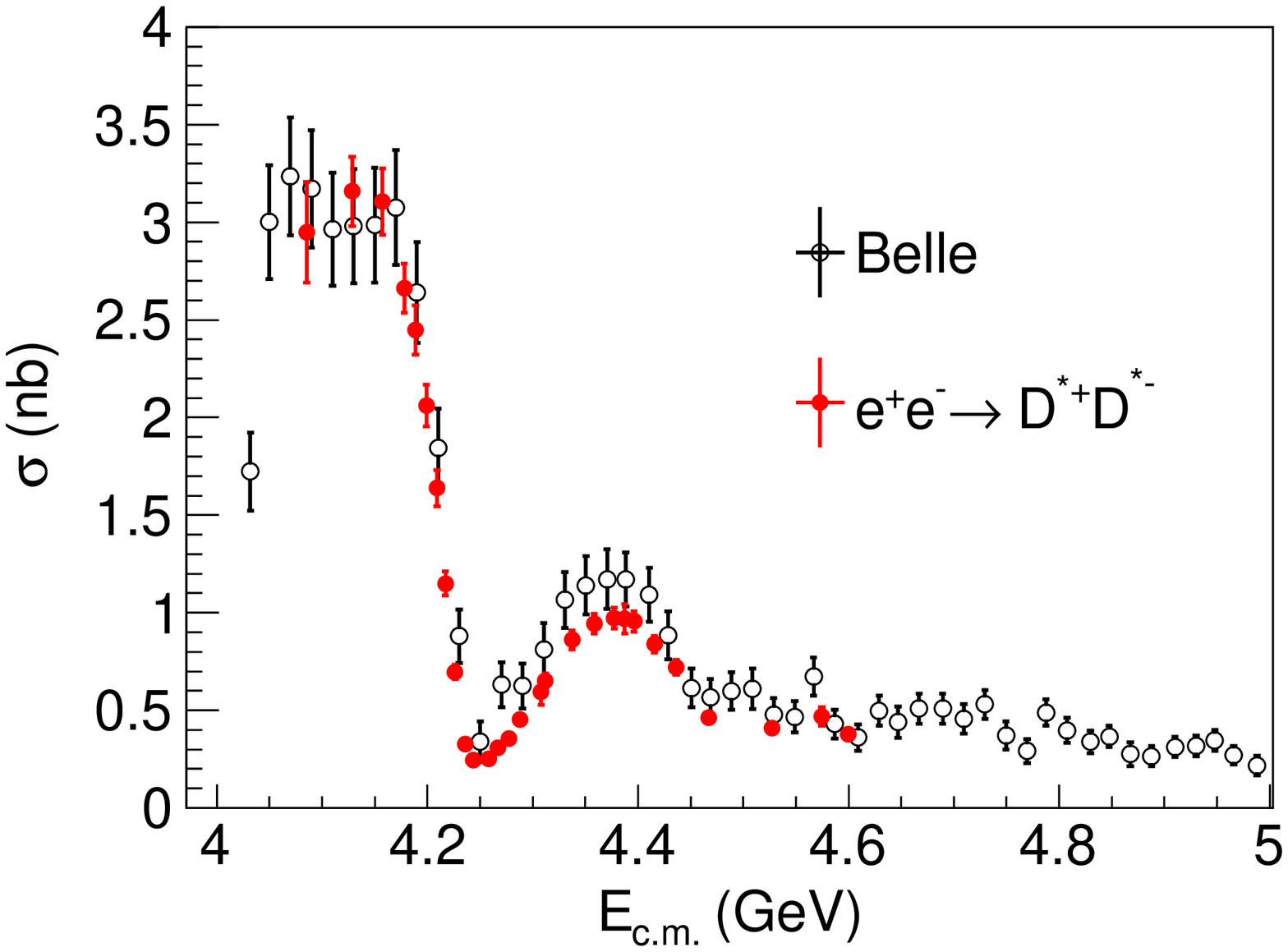}
 \put(-50,110){\textbf{(b)}}
 } 
\end{center}
\caption{ (a) The Born cross sections of $\ee\to\dsds$  as  a function of the c.m.\ energy for the reconstructed $D^{*+}$ (red dots) 
and $D^{*-}$ candidates (blue triangles). (b) The comparison of  the average  cross sections for  $\ee\to\dsds$ between this work (red dots) and those of the Belle experiment~\cite{DstarDstar-belle-2018} (black circles). Error bars are the quadratic sum of statistical and systematic uncertainties.}
\label{fig:born cross section}
\end{figure*}

The cross sections of $\ee\to\dsds$ for different c.m.\ energy points are shown in Figure~\ref{fig:born cross section}(a) for the reconstructed $D^{*+}$ and $D^{*-}$ events, respectively. They are in agreement with each other within uncertainties.
The average cross sections are calculated using the same method as in Ref.~\cite{combine}, where correlations between two measurements  are considered.  
 Figure~\ref{fig:born cross section}(b)  shows the comparison of the average  cross sections of $\ee\to\dsds$ between this work and those of the
Belle~\cite{DstarDstar-belle-2018} experiment.  
 The results are  overall compatible.

\begin{figure*}[htbp]
\begin{center}
 \subfigure{
\label{fig:born cross section b}
\includegraphics[width=0.3\paperwidth]
 {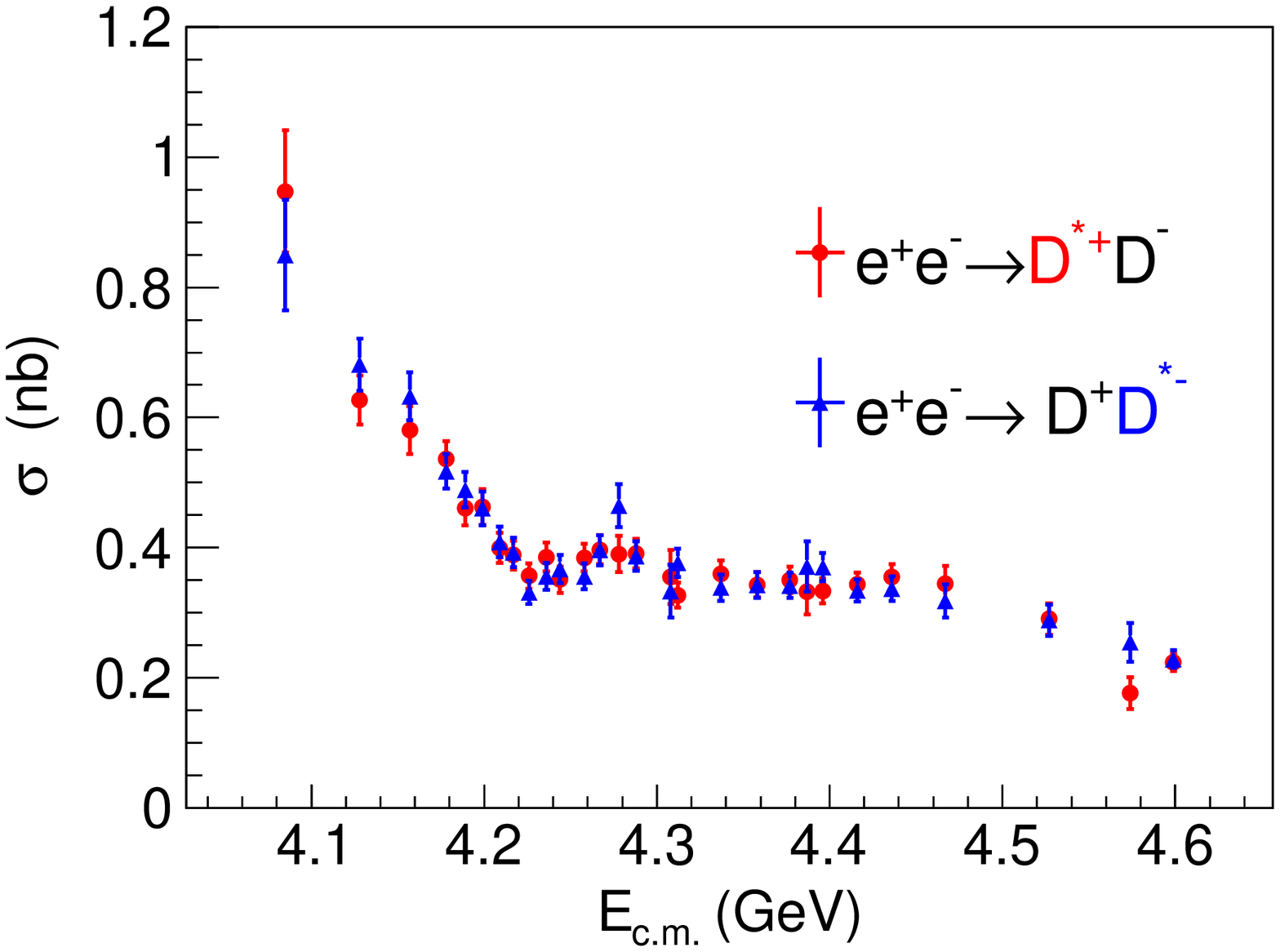}
  \put(-50,110){\textbf{(a)}}
 }
 \subfigure{
\label{fig:born cross section b}
\includegraphics[width=0.3\paperwidth]
 {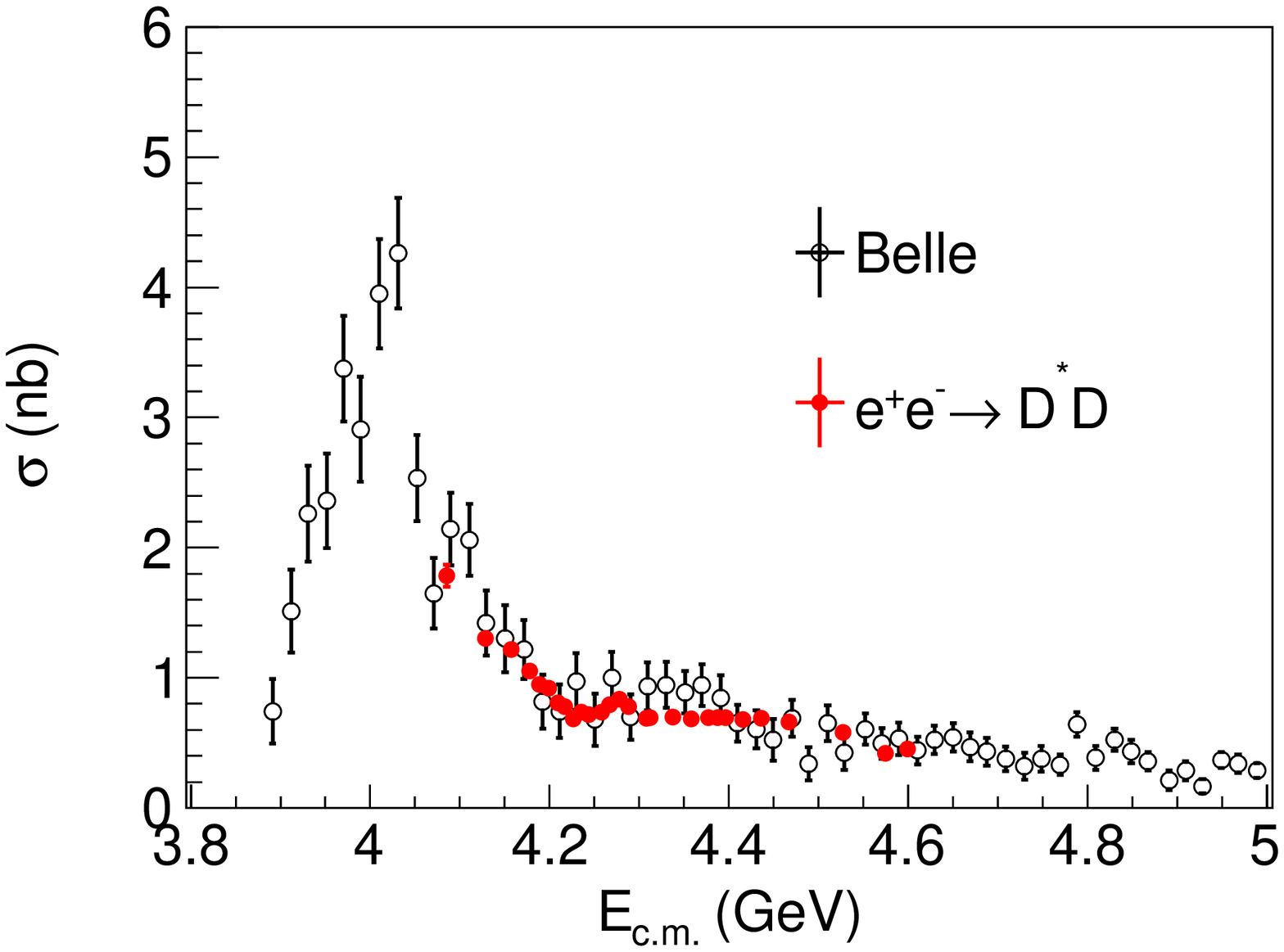}
 \put(-50,110){\textbf{(b)}}
 }
 \end{center}
\caption{ (a) The Born cross sections of the reaction channel  $\ee\to\dsd$ for the reconstructed $D^{*+}$ candidates (red dots)  and  $\ee\to\dds$ for the reconstructed $D^{*-}$ candidates (blue triangles) as  functions of the c.m.\ energy.
(b) The comparison of  the  combined  cross sections for $\ee\to \dsd+c.c.$ between this work (red dots) and those of the Belle experiment~\cite{DstarDstar-belle-2018} (black circles).
Error bars  are the quadratic sum of statistical and systematic uncertainties.}
\label{fig:born cross section2}
\end{figure*}

The cross sections of $\ee\to\dsd$ and $\ee\to\dds$  for different c.m.\ energy points are shown in Figure~\ref{fig:born cross section2}(a). Again, the two results are in good agreement.  
The cross sections of $\ee\to\dsd$ and $\ee\to\dds$ are averaged using the same method as described above. The  total cross sections of $\ee\to\dsd+c.c.$ are twice the average values. The comparison of the cross section of $\ee\to\dsd+c.c.$  between this work and those of  the Belle~\cite{DstarDstar-belle-2018} experiment is shown in Figure~\ref{fig:born cross section2}(b).  They are overall compatible.

\section{Systematic uncertainties}
\label{sec:uncertainty}

The systematic uncertainties in the cross-section measurements mainly come from luminosity determination, 
track reconstruction efficiency,  PID efficiency, and the branching fractions of the charmed meson decays, kinematic fit, ISR correction factor, fit range and modeling of the signal and background shapes. 
The uncertainty from the vacuum polarization is negligible. The uncertainties due to luminosity, 
track reconstruction efficiency,  PID efficiency, and the branching fractions of charmed meson decays are common 
  while the other uncertainties  are individual  and uncommon for reconstructed $\dsp$ and $\dsm$ candidates.

\begin{itemize}
\item \emph{Luminosity, track reconstruction efficiency, and PID efficiency.} 
The integrated luminosity is measured using Bhabha scattering events with an uncertainty of $1.0\%$~\cite{luminosity-measurement, luminosity-measurement2}. 
The uncertainty of the track reconstruction efficiency is 1.0\% per track, taken from Ref.~\cite{intro-Y4220-bes3-open-charm}.  
The uncertainty associated with PID efficiency is taken conservatively to be 1.0\% per track ~\cite{intro-Y4220-bes3-open-charm}. 

\item \emph{Branching fractions.}  
The uncertainties of the branching fractions $\mathcal{B}(\dsp\to\pip D^0)$  and $\mathcal{B}( D^0 \to K^- \pip)$ are 0.74\% and 0.78\%~\cite{pdg}, respectively. 

\end{itemize}

Therefore, the  common systematic uncertainty is 4.49\% in this analysis by summing the individual ones in quadrature.

\begin{itemize}

\item \emph{Kinematic fit.} 
The systematic uncertainty from the kinematic fit is estimated by correcting the helix parameters of charged tracks according to the method described in
Ref.~\cite{bes3-kinematicfit-eff}. The signal MC sample with the track helix parameter correction applied is taken as the nominal one. 
The difference between detection efficiencies obtained from MC samples with and without correction 
is taken as the systematic uncertainty. It is in the range between 0.86\% and 2.40\% for different energy points.

\item \emph{ISR correction factor.} 
The line shape of the cross section affects the ISR correction factor and the reconstruction efficiency. There are three uncertainties involved in the ISR correction factor.
%For the reactions $\ee\to\dsds$ and $\ee\to\dsd$, 
Firstly, the differences of (1+$\delta$)$\epsilon$ between the last two iterations are taken as 
the systematic uncertainties and are in the range between 0.00\% and 2.71\% (ISRI). 
  Secondly, the input cross section at each energy point is changed randomly by varying the central value within its uncertainty and the  $(1+\delta)$ and the efficiency $\epsilon$ are recalculated.
  This process is performed for 1000 times, and a Gaussian function is used to fit the $(1+\delta)\epsilon$ distribution. 
  The width of the  Gaussian function represents the  statistical uncertainty of input line shape and varies between 0.01\% and 0.60\% (ISR2). 
Lastly, an alternative smooth method of LOWESS~~\cite{lowess} is used instead of TSpline3~~\cite{tspline} for the  input cross section line shape, and 
 the differences in  $(1+\delta)\epsilon$ values, varying between  0.01\% to 3.73\% (ISR3), are taken as the  systematic uncertainties.

\item \emph{Fit range.}  
The systematic uncertainty caused by the choice of the fit range is estimated by varying the upper and lower bounds of the fit range by $\pm$ 10 MeV/$c^2$. 
 Toy MC samples are generated using the obtained PDFs and  the corresponding 2D fits are performed in the new fit range. This process has been repeated for 2000 times.  The differences between the input values and the means of the output values are taken as the systematic uncertainties which vary between 0.05\% and 1.61\%.

\item \emph{Background shape.} 
The uncertainty attributed to the background shape is estimated by changing the background shape in $\pip D^0$ recoil mass spectrum 
from the second-order Chebyshev polynomial to a linear function. The uncertainties due to the Argus function and PDFs of the peaking background contributions are  checked and are found to be negligible.  
The systematic uncertainties due to the background shape vary between 0.01\% and 1.95\%.

\item \emph{Signal Shape.} 
The parameters of the Gaussian function, i.e., mean and resolution,  describe the differences
between data and signal MC simulation.  
The systematic uncertainty is assigned by varying the mean and resolution by $\pm$ the corresponding uncertainties when performing the 2D fit.  The systematic uncertainties are in the ranges of 0.01\% to 3.54\% and 0.02\% to 5.01\%, respectively.

\end{itemize}

All the uncommon systematic uncertainties for the reconstructed $\dsp$ and $\dsm$ candidates are summarized in Tables~\ref{table:Unscertainties2}
and~\ref{table:Unscertainties}, respectively. 
The total uncommon systematic uncertainties at each energy point are the sum of individual ones in
quadrature.

% Finally, using the same method as in Ref.~\cite{combine}, the  correlated  systematic uncertainty, the total indepdent systematic 
%uncertainties for the reconstructed $\dsp$ and $\dsm$ candidates, as well as the statistical uncertainties are used to calculate 
%the average cross sections of $\ee\to\dsds$, the combined cross sections of $\ee\to\dsd+c.c.$, and their corresponding uncertainties  as show in Tables~\ref{table:cross-sig} and~\ref{table:cross-bk4}.

\begin{sidewaystable*}[htbp]
%\vspace{8cm}
\begin{center}
\resizebox{22cm}{6.5cm}{
\begin{tabular}{|c|cc|cc|cc|cc|cc|cc|cc|cc|cc|}
\hline
\multirow{2}*{$\sqrt s$~(GeV)}&   \multicolumn{2}{|c|}{Kinematic fit}   &   \multicolumn{2}{|c|}{ISR1}   & \multicolumn{2}{|c|}{ISR2}  & \multicolumn{2}{|c|}{ISR3}   & \multicolumn{2}{|c|}{Fit range}  &  \multicolumn{2}{|c|}{Background shape}   & \multicolumn{2}{|c|}{Mean}   &   \multicolumn{2}{|c|}{Resolution}  &\multicolumn{2}{|c|}{$T_{\rm uncommon}$ } \\
  \cline{2-19}  
         &         $\dsds$    &  $\dsd$              &      $\dsds$    &  $\dsd$              &  $\dsds$    & $\dsd$                &  $\dsds$   &  $\dsd$              &    $\dsds$   &    $\dsd$  &          $\dsds$    &  $\dsd$   &          $\dsds$    &  $\dsd$   &    $\dsds$    &  $\dsd$    &    $\dsds$    &  $\dsd$       \\\hline
   %4.0076 &      -   &    1.54  &    -   &    1.37  &     -   &   1.38   &   -   &    0.28  &    -    &   0.65  &   -   &   0.12   &   -    &   0.19  &   -   &    2.59\\
4.0854 &    2.10  &    2.15  &    0.35  &    1.31  &  0.03  &  0.13  &  0.41  &  0.39 &  0.72  & 0.37 &  0.04  &  0.17  &  1.83  &  1.54 & 2.82  &  2.04   & 4.07 &  3.63 \\
4.1285 &    2.40  &    1.69  &    0.68  &    0.61  &  0.03  &  0.11  &  0.36  &  0.78 &  0.69  & 0.41 &  0.01  &  0.65  &  0.22  &  0.05 & 0.76  &  0.29   & 2.73 &  2.12 \\
4.1574 &    1.96  &    2.07  &    1.10  &    2.13  &  0.01  &  0.07  &  0.80  &  0.02 &  0.88  & 0.24 &  0.03  &  0.42  &  0.17  &  0.48 & 0.24  &  0.66   & 2.56 &  3.12 \\
4.1780 &    0.86  &    1.59  &    0.32  &    0.25  &  0.02  &  0.06  &  0.19  &  0.94 &  0.15  & 0.08 &  0.45  &  0.97  &  0.07  &  0.16 & 0.06  &  0.24   & 1.05 &  2.12 \\
4.1886 &    1.40  &    1.61  &    0.68  &    0.49  &  0.03  &  0.12  &  0.46  &  0.13 &  0.28  & 0.56 &  0.11  &  0.22  &  0.01  &  0.20 & 0.08  &  0.03   & 1.65 &  1.81 \\
4.1989 &    1.45  &    1.79  &    0.46  &    1.19  &  0.04  &  0.11  &  0.24  &  0.56 &  0.35  & 0.13 &  0.37  &  0.85  &  0.27  &  0.11 & 0.39  &  0.12   & 1.69 &  2.39 \\
4.2092 &    2.40  &    1.55  &    1.21  &    1.09  &  0.05  &  0.14  &  0.41  &  0.08 &  0.41  & 0.34 &  0.22  &  0.47  &  0.31  &  0.45 & 0.15  &  0.32   & 2.78 &  2.07 \\
4.2171 &    1.53  &    1.49  &    0.65  &    0.48  &  0.06  &  0.13  &  0.14  &  0.11 &  0.53  & 0.24 &  0.16  &  0.21  &  0.09  &  0.07 & 0.11  &  0.24   & 1.76 &  1.62 \\
4.2263 &    1.50  &    1.79  &    0.39  &    0.27  &  0.09  &  0.12  &  0.12  &  0.25 &  0.48  & 0.26 &  0.10  &  0.11  &  0.32  &  0.07 & 0.25  &  0.13   & 1.68 &  1.86 \\
4.2357 &    1.79  &    1.45  &    0.57  &    0.06  &  0.38  &  0.13  &  2.48  &  0.62 &  1.31  & 0.32 &  0.31  &  0.16  &  0.85  &  0.63 & 1.04  &  0.45   & 3.67 &  1.80 \\
4.2438 &    1.99  &    1.43  &    1.94  &    0.22  &  0.52  &  0.16  &  2.18  &  0.22 &  1.03  & 0.37 &  0.39  &  0.24  &  2.31  &  0.23 & 0.21  &  0.61   & 4.07 &  1.65 \\
4.2580 &    1.96  &    1.69  &    0.99  &    0.05  &  0.30  &  0.10  &  2.89  &  0.09 &  0.88  & 0.22 &  0.14  &  0.13  &  0.25  &  0.12 & 0.78  &  0.12   & 3.84 &  1.82 \\
4.2668 &    2.03  &    1.34  &    1.01  &    0.57  &  0.34  &  0.13  &  0.51  &  0.27 &  1.61  & 0.40 &  0.29  &  0.37  &  0.85  &  0.17 & 1.67  &  0.17   & 3.29 &  1.53 \\
4.2777 &    1.99  &    1.74  &    0.39  &    0.34  &  0.46  &  0.23  &  0.93  &  0.06 &  0.96  & 0.18 &  0.03  &  0.10  &  2.75  &  0.65 & 1.83  &  0.39   & 4.53 &  1.96 \\
4.2879 &    1.82  &    1.39  &    1.91  &    0.38  &  0.15  &  0.14  &  0.09  &  0.29 &  1.45  & 0.61 &  0.25  &  0.17  &  0.46  &  0.34 & 0.33  &  0.94   & 2.46 &  1.86 \\
4.3079 &    1.86  &    1.56  &    0.49  &    0.09  &  0.32  &  0.34  &  3.73  &  1.07 &  0.17  & 0.62 &  0.81  &  1.95  &  1.55  &  0.70 & 2.35  &  0.86   & 5.12 &  3.02 \\
4.3121 &    1.83  &    1.48  &    0.41  &    0.05  &  0.17  &  0.32  &  0.61  &  0.32 &  0.52  & 0.39 &  0.62  &  0.60  &  0.17  &  0.26 & 0.06  &  0.91   & 2.15 &  1.95 \\
4.3374 &    1.76  &    1.41  &    0.43  &    0.10  &  0.06  &  0.13  &  1.28  &  0.20 &  0.66  & 0.76 &  0.37  &  0.25  &  0.15  &  0.30 & 0.03  &  0.30   & 2.32 &  1.70 \\
4.3583 &    1.68  &    1.39  &    0.21  &    0.06  &  0.04  &  0.12  &  0.62  &  0.03 &  0.27  & 0.51 &  0.27  &  0.23  &  0.24  &  0.09 & 0.39  &  0.23   & 2.47 &  1.71 \\
4.3774 &    1.20  &    1.35  &    1.59  &    0.77  &  0.04  &  0.11  &  1.60  &  0.32 &  0.91  & 0.31 &  0.43  &  0.36  &  0.09  &  0.11 & 0.36  &  0.17   & 2.68 &  1.71 \\
4.3874 &    1.22  &    1.44  &    1.43  &    0.85  &  0.15  &  0.54  &  1.43  &  0.12 &  0.31  & 0.38 &  0.31  &  0.52  &  0.31  &  0.08 & 0.32  &  0.11   & 2.84 &  1.72 \\
4.3964 &    1.48  &    1.30  &    2.04  &    0.36  &  0.10  &  0.19  &  0.41  &  0.09 &  0.17  & 0.40 &  0.32  &  0.16  &  0.11  &  0.27 & 0.02  &  0.46   & 1.69 &  1.62 \\
4.4156 &    1.26  &    1.30  &    0.59  &    0.66  &  0.05  &  0.09  &  0.77  &  0.03 &  0.10  & 0.17 &  0.12  &  0.11  &  0.06  &  0.16 & 0.04  &  0.18   & 1.53 &  1.36 \\
4.4362 &    1.53  &    1.68  &    0.37  &    0.26  &  0.07  &  0.10  &  0.15  &  0.14 &  0.70  & 0.20 &  0.48  &  0.36  &  0.10  &  0.09 & 0.28  &  0.17   & 1.94 &  1.75 \\
4.4671 &    1.71  &    1.46  &    0.77  &    0.14  &  0.27  &  0.21  &  1.01  &  0.14 &  0.30  & 0.08 &  0.32  &  0.08  &  0.31  &  0.39 & 1.94  &  0.35   & 2.97 &  1.58 \\
4.5271 &    1.20  &    1.20  &    0.86  &    0.16  &  0.19  &  0.23  &  2.11  &  1.09 &  0.11  & 0.32 &  0.02  &  0.03  &  1.87  &  0.49 & 1.03  &  0.83   & 3.24 &  2.01 \\
4.5745 &    1.30  &    1.32  &    0.18  &    0.56  &  0.17  &  0.60  &  0.93  &  2.27 &  0.11  & 0.97 &  0.57  &  0.59  &  1.08  &  3.54 & 0.92  &  0.30   & 2.31 &  4.66 \\
4.5995 &    1.26  &    1.42  &    0.63  &    0.73  &  0.21  &  0.23  &  0.27  &  0.27 &  0.19  & 0.21 &  0.26  &  0.19  &  0.22  &  0.15 & 0.34  &  0.40   & 1.42 &  1.70 \\\hline

   \end{tabular}
}
\caption{
The uncommon systematic uncertainties from kinematic fit, ISR correction factor, fit range, background shape, mean and resolution of signal shape  for the reconstructed $\dsp$ candidates~(in units of \%). 
The symbol $T_{\rm uncommon}$ is the total uncommon systematic uncertainty for the reconstructed $\dsp$ candidates at each energy point obtained by summing  
individual ones in quadrature. }\label{table:Unscertainties2}
\end{center}
\end{sidewaystable*}

\begin{sidewaystable*}[htbp]
%\vspace{8cm}
\begin{center}
\resizebox{22cm}{6.5cm}{
%\resizebox{16.5cm}{
\begin{tabular}{|c|cc|cc|cc|cc|cc|cc|cc|cc|cc|}
\hline
\multirow{2}*{$\sqrt s$~(GeV)}&   \multicolumn{2}{|c|}{Kinematic fit}   &   \multicolumn{2}{|c|}{ISR1}   &   \multicolumn{2}{|c|}{ISR2}  &   \multicolumn{2}{|c|}{ISR3} & \multicolumn{2}{|c|}{Fit range}  &  \multicolumn{2}{|c|}{Background shape}   & \multicolumn{2}{|c|}{Mean}   &   \multicolumn{2}{|c|}{      Resolution}  &\multicolumn{2}{|c|}{$T_{\rm uncommon}$} \\  
  \cline{2-19}         
         &        $\dsds$    &  $\dds$     &        $\dsds$    &  $\dds$            &    $\dsds$    &  $\dds$              &  $\dsds$    & $\dds$                 &  $\dsds$   &  $\dds$  &    $\dsds$   &    $\dds$  &          $\dsds$    &  $\dds$    &          $\dsds$    &  $\dds$    &     $\dsds$    &  $\dds$         \\\hline
4.0854 & 1.10 & 1.74  & 0.62  & 1.53  & 0.03 &  0.17  &  0.17  &  0.01 & 0.78 &  0.15 &  0.51  & 1.55  &  0.49   & 1.44  &  0.26   & 1.55   & 1.67 &  3.50 \\   
4.1285 & 2.19 & 1.73  & 0.95  & 0.35  & 0.03 &  0.10  &  0.13  &  0.30 & 0.25 &  0.15 &  0.09  & 0.61  &  0.27   & 0.29  &  0.05   & 0.39   & 2.43 &  1.96 \\4.1574 & 2.18 & 1.75  & 0.65  & 0.01  & 0.01 &  0.07  &  0.41  &  0.47 & 0.17 &  0.36 &  0.02  & 0.52  &  0.11   & 0.18  &  0.18   & 0.15   & 2.33 &  1.94 \\   
4.1780 & 1.19 & 1.71  & 1.91  & 0.88  & 0.02 &  0.06  &  0.37  &  0.01 & 0.55 &  0.15 &  0.41  & 0.87  &  0.04   & 0.11  &  0.08   & 0.04   & 2.38 &  2.12 \\4.1886 & 1.87 & 1.49  & 0.05  & 0.18  & 0.03 &  0.11  &  0.14  &  0.10 & 0.93 &  0.05 &  0.45  & 0.86  &  0.30   & 0.15  &  0.09   & 0.24   & 2.17 &  1.76 \\   
4.1989 & 2.04 & 1.17  & 0.31  & 0.35  & 0.04 &  0.11  &  0.25  &  0.27 & 0.72 &  0.35 &  0.21  & 0.45  &  0.12   & 0.38  &  0.35   & 0.12   & 2.24 &  1.44 \\4.2092 & 1.86 & 1.64  & 2.07  & 0.60  & 0.04 &  0.13  &  0.75  &  0.07 & 0.58 &  0.94 &  0.24  & 0.46  &  0.37   & 0.20  &  0.14   & 0.08   & 2.98 &  2.05 \\    
4.2171 & 1.31 & 1.93  & 1.44  & 0.29  & 0.06 &  0.13  &  0.46  &  0.22 & 0.91 &  0.48 &  0.15  & 0.22  &  0.19   & 0.05  &  0.55   & 0.31   & 2.27 &  2.06 \\4.2263 & 1.23 & 1.71  & 0.74  & 0.20  & 0.11 &  0.13  &  0.83  &  0.58 & 0.13 &  0.13 &  0.08  & 0.11  &  0.09   & 0.12  &  0.32   & 0.15   & 1.70 &  1.84 \\  
4.2357 & 2.16 & 1.82  & 2.08  & 0.12  & 0.36 &  0.14  &  1.64  &  0.06 & 1.06 &  0.22 &  0.08  & 0.04  &  0.72   & 0.22  &  0.36   & 0.15   & 3.69 &  1.87 \\4.2438 & 1.81 & 1.75  & 2.24  & 1.11  & 0.51 &  0.14  &  1.46  &  0.30 & 1.20 &  0.37 &  0.27  & 0.14  &  0.60   & 0.33  &  2.28   & 0.12   & 4.21 &  2.17 \\ 
4.2580 & 1.92 & 1.69  & 2.71  & 1.15  & 0.33 &  0.11  &  1.02  &  0.53 & 0.84 &  0.18 &  0.13  & 0.15  &  0.49   & 0.11  &  0.23   & 0.16   & 3.63 &  2.14 \\4.2668 & 1.76 & 1.88  & 0.34  & 0.64  & 0.31 &  0.13  &  0.09  &  0.15 & 0.21 &  0.19 &  0.02  & 0.13  &  0.27   & 0.40  &  0.77   & 0.43   & 2.01 &  2.09 \\ 
4.2777 & 1.88 & 1.70  & 0.12  & 1.06  & 0.35 &  0.18  &  0.22  &  1.19 & 0.89 &  0.14 &  0.38  & 0.22  &  1.16   & 0.69  &  0.88   & 0.18   & 2.60 &  2.46 \\4.2879 & 2.06 & 1.47  & 1.02  & 0.41  & 0.15 &  0.15  &  0.45  &  0.05 & 0.26 &  0.11 &  0.21  & 0.31  &  0.72   & 0.59  &  0.67   & 0.25   & 2.57 &  1.70 \\ 
4.3079 & 1.88 & 1.80  & 0.72  & 0.00  & 0.18 &  0.36  &  2.92  &  2.76 & 0.21 &  0.52 &  0.34  & 0.14  &  2.40   & 1.30  &  0.10   & 1.69   & 4.30 &  3.98 \\4.3121 & 1.18 & 1.65  & 0.03  & 0.41  & 0.22 &  0.26  &  0.60  &  0.29 & 0.35 &  0.31 &  0.26  & 0.28  &  0.24   & 0.02  &  0.41   & 0.14   & 1.49 &  1.80 \\ 
4.3374 & 1.90 & 1.58  & 0.12  & 1.49  & 0.06 &  0.15  &  0.67  &  0.77 & 0.28 &  0.44 &  0.19  & 0.10  &  0.17   & 0.35  &  0.30   & 0.35   & 2.07 &  2.41 \\
4.3583 & 1.45 & 1.71  & 0.13  & 0.90  & 0.05 &  0.12  &  0.02  &  0.07 & 0.27 &  0.61 &  0.26  & 0.25  &  0.06   & 0.12  &  0.20   & 0.37   & 1.52 &  2.08 \\ 
4.3774 & 1.53 & 1.26  & 1.05  & 0.03  & 0.04 &  0.11  &  1.06  &  0.40 & 0.76 &  0.38 &  0.47  & 0.48  &  0.19   & 0.30  &  0.15   & 0.27   & 2.33 &  1.52 \\
4.3874 & 1.26 & 1.76  & 0.38  & 0.38  & 0.24 &  0.54  &  0.75  &  0.22 & 0.08 &  0.05 &  0.98  & 0.59  &  0.71   & 0.46  &  0.92   & 1.47   & 2.16 &  2.51 \\ 
4.3964 & 1.78 & 1.62  & 1.86  & 1.63  & 0.08 &  0.18  &  0.51  &  0.26 & 0.26 &  0.49 &  0.25  & 0.23  &  0.24   & 0.21  &  0.28   & 0.22   & 2.67 &  2.40 \\
4.4156 & 1.58 & 1.36  & 0.59  & 0.21  & 0.05 &  0.10  &  0.42  &  0.22 & 0.19 &  0.07 &  0.19  & 0.10  &  0.19   & 0.08  &  0.23   & 0.10   & 1.78 &  1.41 \\ 
4.4362 & 1.07 & 1.64  & 1.02  & 0.34  & 0.06 &  0.11  &  0.35  &  0.02 & 0.18 &  0.37 &  0.28  & 0.19  &  0.13   & 0.02  &  0.32   & 0.13   & 1.60 &  1.74 \\
4.4671 & 1.57 & 1.44  & 0.22  & 0.01  & 0.26 &  0.22  &  0.50  &  0.09 & 0.09 &  0.22 &  0.21  & 0.11  &  0.78   & 0.80  &  0.81   & 0.42   & 2.04 &  1.73 \\ 
4.5271 & 1.48 & 1.52  & 0.17  & 1.07  & 0.16 &  0.21  &  2.15  &  0.03 & 0.11 &  0.46 &  0.13  & 0.14  &  1.19   & 0.86  &  1.39   & 0.09   & 3.20 &  2.12 \\
4.5745 & 1.52 & 1.37  & 0.03  & 1.25  & 0.18 &  0.36  &  0.56  &  0.15 & 0.61 &  1.01 &  0.36  & 0.09  &  0.12   & 0.73  &  5.01   & 2.28   & 5.31 &  3.22 \\ 4.5995 & 1.34 & 1.60  & 0.22  & 1.64  & 0.18 &  0.26  &  0.12  &  0.01 & 0.16 &  0.08 &  0.35  & 0.09  &  0.44   & 0.02  &  0.73   & 0.38   & 1.66 &  2.33 \\\hline

    \end{tabular}
}
\caption{The uncommon systematic uncertainties from kinematic fit, ISR correction factor, fit range, background shape, mean and resolution of signal shape  for the reconstructed $\dsm$ candidates~(in units of \%).
The symbol $T_{\rm uncommon}$ is the total uncommon systematic uncertainty  for the reconstructed $\dsm$ candidates at each energy point obtained 
by summing individual ones in quadrature. }\label{table:Unscertainties}
\end{center}
\end{sidewaystable*}

\section{Summary}
\label{sec:summ}

A measurement of the cross sections of the $\ee\to\dsds$ 
and $\ee\to\dsd$ processes is presented using 28 data samples corresponding to a total integrated luminosity of $15.7~{\rm fb}^{-1}$ with c.m.\ energies between 4.085 and 4.600~GeV. 
The  cross sections  are consistent with and  more precise than those of the Belle,  BaBar, and CLEO experiments.  
The structures in the cross sections measured by the previous experiments are confirmed.
In order to finally reveal the nature of the vector charmonium(-like) states in this energy region, measurements on other two-body 
open-charm modes such as $D\bar{D}$, $D_s^+D_s^-$, $D_s^+D_s^{*-}+c.c.$, $D_s^{*+}D_s^{*-}$~\cite{charmonium1980} and multi-body modes such as 
$\pi D\bar{D}$, $\pi D^*\bar{D}+c.c.$, $\pi \pi D\bar{D}$ are necessary. 
More sophisticated models with further studies on coupled-channel effect are also needed.

\acknowledgments

The BESIII collaboration thanks the staff of BEPCII and the IHEP computing center for their strong support. This work is supported in part by National Key R\&D Program of China under Contracts Nos. 2020YFA0406300, 2020YFA0406400; National Natural Science Foundation of China (NSFC) under Contracts Nos. 11625523, 11635010, 11735014, 11822506, 11835012, 11935015, 11935016, 11935018, 11961141012, 12022510, 12025502, 12035009, 12035013, 12061131003; the Chinese Academy of Sciences (CAS) Large-Scale Scientific Facility Program; Joint Large-Scale Scientific Facility Funds of the NSFC and CAS under Contracts Nos. U1732263, U1832207; CAS Key Research Program of Frontier Sciences under Contract No. QYZDJ-SSW-SLH040; 100 Talents Program of CAS; INPAC and Shanghai Key Laboratory for Particle Physics and Cosmology; ERC under Contract No. 758462; European Union Horizon 2020 research and innovation programme under Contract No. Marie Sklodowska-Curie grant agreement No 894790; German Research Foundation DFG under Contracts Nos. 443159800, Collaborative Research Center CRC 1044, FOR 2359, GRK 214; Istituto Nazionale di Fisica Nucleare, Italy; Ministry of Development of Turkey under Contract No. DPT2006K-120470; National Science and Technology fund; Olle Engkvist Foundation under Contract No. 200-0605; STFC (United Kingdom); The Knut and Alice Wallenberg Foundation (Sweden) under Contract No. 2016.0157; The Royal Society, UK under Contracts Nos. DH140054, DH160214; The Swedish Research Council; U. S. Department of Energy under Contracts Nos. DE-FG02-05ER41374, DE-SC-0012069.

%\bibliographystyle{JHEP.bst}

%\newpage
%\label{sec:uncertainty}
%\bes
%\input{author.tex}


\begin{thebibliography}{99}



\bibitem{barnes} T. Barnes, S. Godfrey, and E. S. Swanson, \emph{Higher charmonia}, 
 \href{https://journals.aps.org/prd/abstract/10.1103/PhysRevD.72.054026}{\emph{Phys.\ Rev.\ }{\bf D72} (2005) 054026}.


\bibitem{intro-BaBar-Y4260} BaBar Collaboration, \emph{Observation of a broad structure in the $\ppjpsi$ mass spectrum around $4.26$ GeV/$c^2$},
 \href{https://journals.aps.org/prl/abstract/10.1103/PhysRevLett.95.142001}{\emph{Phys.\ Rev.\ Lett.\ }{\bf 95} (2005) 142001}.

\bibitem{intro-BaBar-Y4260-2012} BaBar Collaboration, \emph{Study of the reaction $\ee\to\jpsi\pp$ via initial-state radiation at BaBar},
 \href{https://journals.aps.org/prd/abstract/10.1103/PhysRevD.86.051102}{\emph{Phys.\ Rev.\ }{\bf D86} (2012) 051102(R)}.

\bibitem{intro-Belle-Y4260}  Belle Collaboration, \emph{Measurement of the $\ee\to\ppjpsi$ cross section via initial-state radiation at Belle}, 
 \href{https://journals.aps.org/prl/abstract/10.1103/PhysRevLett.99.182004}{\emph{Phys.\ Rev.\ Lett.\ }{\bf 99} (2007) 182004}.

\bibitem{intro-Belle-Y4260-2} Belle Collaboration, \emph{Study of $\ee\to\ppjpsi$ and observation of a charged charmoniumlike state at Belle},
 \href{https://journals.aps.org/prl/abstract/10.1103/PhysRevLett.110.252002}{\emph{Phys.\ Rev.\ Lett.\ }{\bf 110} (2013) 252002}.

\bibitem{intro-BaBar-Y4360} BaBar Collaboration, \emph{Evidence of a broad structure at an invariant mass of $4.32$  GeV/$c^2$ in the reaction $\ee\to\pppsip$ measured at BaBar},
 \href{https://journals.aps.org/prl/abstract/10.1103/PhysRevLett.98.212001}{\emph{Phys.\ Rev.\ Lett.\ }{\bf 98} (2007) 212001}.

\bibitem{intro-Belle-Y4360-Y4660} Belle Collaboration, \emph{Observation of two resonant structures in $\ee\to\pppsip$ via initial-state radiation at Belle},
 \href{https://journals.aps.org/prl/abstract/10.1103/PhysRevLett.99.142002}{\emph{Phys.\ Rev.\ Lett.\ }{\bf 99} (2007) 142002}.

\bibitem{intro-BaBar-Y4360-Y4660-2014} BaBar Collaboration, \emph{Study of the reaction $\ee\to\psip\pp$ via initial-state radiation at BaBar}, 
   \href{https://journals.aps.org/prd/abstract/10.1103/PhysRevD.89.111103}{\emph{Phys.\ Rev.\ }{\bf D89} (2014) 111103(R)}.

\bibitem{ABLIKIM-2017B} BESIII Collaboration, \emph{Precise measurement of the $e^+e^-\to \pi^+\pi^-J/\psi$ cross section at center-of-mass energies from $3.77$ to $4.60$ GeV}, \href{https://journals.aps.org/prl/abstract/10.1103/PhysRevLett.118.092001}{\emph{Phys.\ Rev.\ Lett.\ }{\bf 118} (2017) 092001}




\bibitem{bes3-Y4230-pipihc} BESIII Collaboration, \emph{Evidence of two resonant structures in $\ee\to\pphc$}, 
 \href{https://journals.aps.org/prl/abstract/10.1103/PhysRevLett.118.092002}{ \emph{Phys.\ Rev.\ Lett.\ }{\bf 118} (2017) 092002}.

\bibitem{bes3-Y4230-omegachic0} BESIII Collaboration, \emph{Cross section measurements of $\ee\to\omega\chi_{c0}$  from $\sqrt{s}=4.178$ to $4.278$ GeV}, 
 \href{https://journals.aps.org/prd/abstract/10.1103/PhysRevD.99.091103}{\emph{Phys.\ Rev.\ }{\bf D99} (2019) 091103(R)}.

   \bibitem{intro-Y4220-bes3-open-charm} BESIII Collaboration, \emph{Evidence of a resonant structure in the $\ee\to\pip\dz\dsm$ cross section between $4.05$ and $4.60$ GeV}, 
 \href{https://journals.aps.org/prl/abstract/10.1103/PhysRevLett.122.102002}{\emph{Phys.\ Rev.\ Lett.\ }{\bf 122} (2019) 102002}.

\bibitem{theory-Y-states-chenhuaxing-2016} H. X. Chen, W. Chen, X. Liu, and S. L. Zhu, \emph{The hidden-charm pentaquark and tetraquark states}, 
 \href{https://www.sciencedirect.com/science/article/pii/S037015731630103X?via%3Dihub}{\emph{Phys.\ Rep.\ }{\bf 639} (2016) 1}.

\bibitem{theory-Y-states-Esposito-2017} A. Esposito, A. Pilloni and A. D. Polosa, \emph{Multiquark resonances}, 
 \href{https://www.sciencedirect.com/science/article/pii/S037015731630391X?via%3Dihub}{\emph{Phys.\ Rep.\ }{\bf 668} (2017) 1}.

\bibitem{theory-Y-states-Richard-2017} R. F. Lebed, R. E. Mitchell, and E. S.  Swanson, \emph{Heavy-quark QCD exotica}, 
 \href{https://linkinghub.elsevier.com/retrieve/pii/S0146641016300734}{\emph{Prog.\ Part.\ Nucl.\ Phys.\ }{\bf 93} (2017) 143}.

\bibitem{theory-Y-states-Ali-2017} A. Ali, J. S. Lange and S. Stone, \emph{Exotics: Heavy pentaquarks and tetraquarks}, 
 \href{https://www.sciencedirect.com/science/article/pii/S0146641017300716?via%3Dihub}{\emph{Prog.\ Part.\ Nucl.\ Phys.\ }{\bf 97} (2017) 123}.

\bibitem{theory-Y-states-Stephen-2018} S. L. Olsen, T. Skwarnicki, and D. Zieminska, \emph{Nonstandard heavy mesons and baryons: Experimental evidence}, 
 \href{https://journals.aps.org/rmp/abstract/10.1103/RevModPhys.90.015003}{\emph{Rev.\ Mod.\ Phys.\ }{\bf 90} (2018) 015003}.

\bibitem{theory-Y-states-guofenghun-2018} F. K. Guo, C. Hanhart, U. G. Mei\ss ner, Q. Wang, Q. Zhao, and B. S. Zou, \emph{Hadronic molecules}, 
 \href{https://journals.aps.org/rmp/abstract/10.1103/RevModPhys.90.015004}{\emph{Rev.\ Mod.\ Phys.\ }{\bf 90} (2018) 015004}.

\bibitem{theory-Y-states-Brambilla-2020} N. Brambilla, S. Eidelman, C. Hanhart, A. Nefediev,
C. P. Shen, C. E. Thomas, A. Vairo, and C. Z. Yuan, \emph{The XYZ states: Experimental and theoretical status and perspectives}, 
 \href{https://www.sciencedirect.com/science/article/pii/S0370157320301915}{\emph{Phys.\ Rep.\ }{\bf 873} (2020) 1}.



 %\bibitem{intro-Y4630-belle-open-charm} Belle Collaboration, \emph{Measurement of the near-threshold $\ee\to D^{(*)\pm}D^{(*)\mp}$ cross section using initial-state radiation}, 
  %  \href{https://journals.aps.org/prd/abstract/10.1103/PhysRevD.100.111103}{\emph{Phys.\ Rev.\ }{\bf D100} (2019) 111103(R)}.
 
  %\bibitem{Y4630-belle-2020} Belle Collaboration, \emph{Evidence for a vector charmoniumlike state in $\ee\to D^{+}_{s}D^{*}_{s2}(2573)^{-}+c.c.$}, 
  % \href{https://journals.aps.org/prd/pdf/10.1103/PhysRevD.101.091101}{\emph{Phys.\ Rev.\ }{\bf D101} (2020) 091101(R)}.

  \bibitem{DstarDstar-belle-2007} Belle Collaboration, \emph{Measurement of the near-threshold $\ee\to D{(*)\pm}D^{(*)\mp}$ cross section using initial-state radiation}, 
 \href{https://journals.aps.org/prl/abstract/10.1103/PhysRevLett.98.092001}{\emph{Phys.\ Lett.\ }{\bf 98} (2007) 092001}.
 

  \bibitem{DstarDstar-belle-2018} Belle Collaboration, \emph{Angular analysis of the $\ee\to D{(*)\pm}D^{(*)\mp}$ process near the open charm threshold using initial-state radiation}, 
  \href{https://journals.aps.org/prd/abstract/10.1103/PhysRevD.97.012002}{\emph{Phys.\ Rev.\ }{\bf D97} (2018) 012002(R)}.

  \bibitem{DstarDstar-babar-2009} BaBar Collaboration, \emph{Exclusive initial-state-radiation production of the $D\bar{D}$, $D^{*}\bar{D}$, and  $D^{*}\bar{D}^{*}$ systems}, 
 \href{https://journals.aps.org/prd/abstract/10.1103/PhysRevD.79.092001}{\emph{Phys.\ Rev.\ }{\bf D79} (2009) 092001}.

   \bibitem{DstarDstar-cleo-2009} CLEO Collaboration, \emph{Measurement of charm production cross sections in  $\ee$ annihilation at energies between $3.97$ and $4.26$ GeV}, 
    \href{https://journals.aps.org/prd/abstract/10.1103/PhysRevD.80.072001}{\emph{Phys.\ Rev.\ }{\bf D80} (2009) 072001}.

   \bibitem{zhaoqiang2018} S.\ R.\ Xue, H.\ J.\ Jing, F.\ K.\ Guo, and Q.\ Zhao,  \emph{Disentangling the role of the $Y(4260)$ in $\ee\to D^{*}\bar{D}^{*}$ and $D^{*}_{s}\bar{D}^{*}_{s}$ via
line shape studies}, 
    \href{https://www.sciencedirect.com/science/article/pii/S0370269318301291?via%3Dihub}{\emph{Phys.\  Lett.\ }{\bf B779} (2018) 402}.


  \bibitem{Uglov} T.\ V.\ Uglov, Y.\ S.\ Kalashnikova, A.\ V.\  Nefediev, G.\ V.\ Pakhlova, and P.\ N.\ Pakhlov, \emph{Exclusive open-charm near-threshold cross sections in a coupled-channel approach}, 
    \href{https://link.springer.com/article/10.1134%2FS0021364017010064}{\emph{JETP Lett.\ }{\bf 105} (2017) 1}.
    
      \bibitem{caoqinfang} Q.\ F.\ Cao, H.\ R.\ Qi, G.\ Y.\ Tang, Y.\ F.\ Xue, and H.\ Q.\ Zheng, \emph{On leptonic width of $X(4260)$}, 
    \href{https://link.springer.com/article/10.1140%2Fepjc%2Fs10052-020-08813-y}{\emph{Eur.\ Phys.\ J.\ }{\bf C81} (2021) 83}.   

\bibitem{Ablikim-2009aa}
  BESIII Collaboration, \emph{Design and construction of the BESIII detector}, 
   \href{https://www.sciencedirect.com/science/article/pii/S0168900209023870?via\%3Dihub}{\emph{Nucl.\ Instrum.\ Meth.\ }{\bf A614} (2010) 345}.
  
\bibitem{bes3-energy-measurement}
BESIII Collaboration, \emph{Measurements of the center-of-mass energies at BESIII via the di-muon process}, 
  \href{http://hepnp.ihep.ac.cn/article/doi/10.1088/1674-1137/40/6/063001}{\emph{Chin.\ Phys.\ }{\bf C40} (2016) 063001}.

\bibitem{luminosity-measurement}
BESIII Collaboration, \emph{Precision measurement of the integrated luminosity of the data taken by BESIII at center-of-mass energies between $3.810$ GeV and $4.600$ GeV}, 
 \href{http://hepnp.ihep.ac.cn/article/doi/10.1088/1674-1137/39/9/093001}{\emph{Chin.\ Phys.\ }{\bf C39} (2015) 093001}. 

\bibitem{luminosity-measurement2}
BESIII Collaboration,  \emph{Luminosity measurements for the R scan experiment at BESIII},
  \href{http://hepnp.ihep.ac.cn/en/article/doi/10.1088/1674-1137/41/6/063001}{\emph{Chin.\ Phys.\ }{\bf C41} (2017) 063001}.

%\bibitem{Ablikim:2009aa}
 % BESIII Collaboration,  \emph{Design and construction of the BESIII detector}, 
  % \href{https://www.sciencedirect.com/science/article/pii/S0168900209023870?via%3Dihub}{\emph{Nucl.\ Instrum.\ Meth.\ }{\bf A 614} (2010) 345}.

\bibitem{Yu-IPAC2016-TUYA01}
   C.~H.~Yu {\it et al.}, \emph{BEPCII performance and beam dynamics studies on luminosity},  
   \href{http://accelconf.web.cern.ch/ipac2016/doi/JACoW-IPAC2016-TUYA01.html}{ in \emph{Proceedings of IPAC $2016$}, Busan, Korea, 2016}.

   
  \bibitem{Ablikim:2019hff}
 BESIII Collaboration, \emph{Future physics programme of BESIII}, 
 \href{http://cpc.ihep.ac.cn/article/doi/10.1088/1674-1137/44/4/040001}{\emph{Chin.\ Phys.\ }{\bf C44} (2020) 040001}.


\bibitem{etof}
 X.~Li {\it et al.}, \emph{Study of MRPC technology for BESIII endcap-TOF upgrade}, 
  \href{https://link.springer.com/article/10.1007/s41605-017-0014-2}{ \emph{Rad.\ Det.\ Tech.\ Meth.\ }{\bf 1} (2017) 13}.

  \bibitem{etof-2}
 Y.~X.~Guo {\it et al.}, \emph{The study of time calibration for upgraded end cap TOF of BESIII}, 
  \href{https://link.springer.com/article/10.1007\%2Fs41605-017-0012-4}{ \emph{Rad.\ Det.\ Tech.\ Meth.\ }{\bf 1} (2017) 15}.

   \bibitem{etof-3}
 P.~Cao {\it et al.}, \emph{Design and construction of the new BESIII endcap Time-of-Flight system with MRPC technology}, 
  \href{https://www.sciencedirect.com/science/article/pii/S0168900219314068}{ \emph{Nucl.\ Instrum.\ Meth.\ }{\bf A953} (2020) 163053}.

\bibitem{geant4} GEANT4 Collaboration, \emph{Geant4—a simulation toolkit}, 
 \href{https://www.sciencedirect.com/science/article/pii/S0168900203013688?via\%3Dihub}{\emph{Nucl.\ Instrum.\ Meth.\ }{\bf A506} (2003) 250}.



\bibitem{ref-kkmc}
  S.~Jadach, B.~F.~L.~Ward, and Z.~Was, \emph{The precision Monte Carlo event generator KK  for two-fermion final states in  $\ee$  collisions}, \href{https://www.sciencedirect.com/science/article/pii/S0010465500000485?via\%3Dihub}{\emph{Comput.\ Phys.\ Commun.\ }{\bf 130} (2000) 260}.

  \bibitem{ref-kkmc2}
  S.~Jadach, B.~F.~L.~Ward, and Z.~Was, \emph{Coherent exclusive exponentiation for precision Monte Carlo calculations}, 
  \href{https://journals.aps.org/prd/abstract/10.1103/PhysRevD.63.113009}{\emph{Phys.\ Rev.\ } {\bf D63} (2001) 113009}.



  
\bibitem{ref-evtgen}
  D.~J.~Lange, \emph{The EvtGen particle decay simulation package}, 
  \href{https://www.sciencedirect.com/science/article/pii/S0168900201000894?via\%3Dihub}{ \emph{Nucl.\ Instrum.\ Meth.\ }{\bf A462} (2001) 152}.

\bibitem{ref-evtgen2}
      R.~G.~Ping, \emph{Event generators at BESIII}, 
  \href{https://iopscience.iop.org/article/10.1088/1674-1137/32/8/001}{\emph{ Chin.\ Phys.\ }{\bf C32} (2008) 599}.

   \bibitem{combine}
  G.~D'Agostini, \emph{On the use of the covariance matrix to fit correlated data}, 
   \href{https://www.sciencedirect.com/science/article/abs/pii/0168900294907196?via%3Dihub}{\emph{Nucl.\ Instrum.\ Meth.\ }{\bf A346} (1994) 306}.

\bibitem{pdg} Particle Data Group, \emph{Review of particle physics}, 
\href{https://academic.oup.com/ptep/article/2020/8/083C01/5891211}{\emph{Prog.\ Theor.\ Exp.\ Phys.\ }{\bf 2020} (2020) 083C01}.

\bibitem{ref:lundcharm}
 J.\ C.\ Chen, G.\ S.\ Huang, X.\ R.\ Qi, D.\ H.\ Zhang, and Y.\ S.\ Zhu, \emph{Event generator for $\jpsi$ and $\psip$ decay}, 
  \href{https://journals.aps.org/prd/abstract/10.1103/PhysRevD.62.034003}{\emph{Phys.\ Rev.\ }{\bf D62} (2000) 034003}.

\bibitem{ref:lundcharm2}
 R.\ L.\ Yang, R.\ G.\
Ping, and H.\ Chen, \emph{Tuning and validation of the Lundcharm model with $\jpsi$ decays}, \href{https://iopscience.iop.org/article/10.1088/0256-307X/31/6/061301}{\emph{Chin.\ Phys.\ Lett.\ }{\bf 31} (2014) 061301}.

  \bibitem{photos}
  E.~Richter-Was, \emph{QED bremsstrahlung in semileptonic $B$ and leptonic $\tau$ decays}, 
   \href{https://www.sciencedirect.com/science/article/abs/pii/037026939390062M?via\%3Dihub}{\emph{Phys.\ Lett.\ } {\bf B303} (1993) 163}.

%\bibitem{topo} 
%X.\ Y.\ Zhou, S.\ X.\ Du, G.\ Li and C.\ P.\ Shen, \emph{TopoAna: A generic tool for the event type analysis of inclusive Monte-Carlo samples in high energy physics experiments}, \href{https://doi.org/10.1016/j.cpc.2020.107540}{\emph{Comput.\ Phys.\ Commun.\ }{\bf 258} (2021) 107540}.



\bibitem{argus} ARGUS Collaboration, \emph{Search for Hadronic $b \to u$ Decays}, 
 \href{https://www.sciencedirect.com/science/article/abs/pii/037026939091293K?via%3Dihub}{\emph{Phys.\ Lett.\ }{\bf B241} (1990) 278}.

   \bibitem{vacuum-polarization-factor}
  S.~Actis {\it et al.} (Working group on radiative corrections and Monte Carlo generators for low energies), \emph{Quest for precision in hadronic cross sections at low energy: Monte Carlo tools vs.\ experimental data}, 
   \href{https://link.springer.com/article/10.1140\%2Fepjc\%2Fs10052-010-1251-4}{\emph{ Eur.\ Phys.\ J.\ }{\bf C66} (2010) 585}.

\bibitem{isr-calculate2}
E.\ A.\ Kuraev and V.\ S.\ Fadin, \emph{Yad.\  Fiz.\ }{\bf 41} (1985) 733  [\emph{Sov.\ J.\  Nucl.\ Phys.\ }{\bf 41} (1985) 466].
 



\bibitem{bes3-kinematicfit-eff}
BESIII Collaboration, \emph{Search for hadronic transition $\chicj\to\etac\pp$ and observation of  $\chicj\to K\bar{K}\pi\pi\pi$}, 
 \href{https://journals.aps.org/prd/abstract/10.1103/PhysRevD.87.012002}{\emph{Phys.\ Rev.\ }{\bf D87} (2013) 012002}.

% \bibitem{intro-huyu-bes3-open-charm} BESIII Collaboration, \emph{Observation of $\ee\to\pp\psi(3770)$ and $D_{1}(2420)^{0}\bar{D}^{0}+c.c.$}, 
% \href{https://journals.aps.org/prd/abstract/10.1103/PhysRevD.100.032005}{\emph{Phys.\ Rev.\ }{\bf D 100} (2019) 032005}.

 %\bibitem{intro-landiao-bes3-open-charm} BESIII Collaboration, \emph{Study of $\ee\dplus\dm\pp$ at center-of-mass energies from $4.36$ to $4.60$ GeV}, 
 %\href{https://www.sciencedirect.com/science/article/pii/S0370269320301994?via%3Dihub}{\emph{Phys.\  Lett.\ }{\bf B 804} (2020) 135395}.


%\bibitem{intro-CLEO-etajpsi} CLEO Collaboration, \emph{Charmonium Decays of   $Y(4260)$, $\psi(4160)$, and $\psi(4040)$}, 
 % \href{https://journals.aps.org/prl/abstract/10.1103/PhysRevLett.96.162003}{\emph{Phys.\ Rev.\ Lett.\ }{\bf 96} (2006) 162003}.

%  \bibitem{DD-belle-2008} Belle Collaboration, \emph{Measurement of the near-threshold $\ee\to D\bar{D}$ cross section using initial-state radiation}, 
% \href{https://journals.aps.org/prd/abstract/10.1103/PhysRevD.77.011103}{\emph{Phys.\ Rev.\ }{\bf D 77} (2008) 011103(R)}.

%\bibitem{D0Dmpip-belle-2008} Belle Collaboration, \emph{Observation of the $\psi(4415)\to D\bar{D}^{*}_{2}(2460)$  decay using initial-state radiation}, 
 %\href{https://journals.aps.org/prl/abstract/10.1103/PhysRevLett.100.062001}{\emph{Phys.\ Rev.\ Lett.\ }{\bf 100} (2008) 062001}.
   
%\bibitem{D0Dmpip-belle-2008} Belle Collaboration, \emph{Observation of the $\psi(4415)\to D\bar{D}^{*}_{2}(2460)$  decay using initial-state radiation}, 
 %\href{https://journals.aps.org/prl/abstract/10.1103/PhysRevLett.100.062001}{\emph{Phys.\ Rev.\ Lett.\ }{\bf 100} (2008) 062001}. 


   %\bibitem{wenyu} W.~Y.~Sun, T.~Liu, M.~Q.~Jing, L.~L.~Wang, B.~Zhong, and W.~M.~Song, \emph{An iterative weighting method to apply ISR correction to $\ee$ hadronic cross-section measurements}, \href{https://journal.hep.com.cn/fop/EN/10.1007/s11467-021-1085-6}{\emph{Front.\ Phys.\ }{\bf 16} (2021) 64501}.



   \bibitem{lowess}
    W.\ S.\ Cleveland, \emph{Robust locally weighted regression and smoothing scatterplots}, 
   \emph{JASA }{\bf 74} (1979) 829. 

    \bibitem{tspline}  
    C.\ H.\ Reinsch, \emph{Smoothing by spline functions}, 
   \emph{Numer.\ Math.\ }{\bf 10} (1967) 177. 
   %\href{https://root.cern.ch/doc/master/classTSpline3.html#a397e96cc4c7a7cbc0cb123826fdebad5}{ \emph{root class}}.

   \bibitem{charmonium1980} 
   E.\ Eichten, K.\ Gottfried, T.\ Kinoshita, K.\ D.\ Lane, and T.\ M.\ Yan, \emph{Charmonium: comparison with experiment}, \href{https://journals.aps.org/prd/abstract/10.1103/PhysRevD.21.203}{\emph{Phys.\ Rev.\ }{\bf D21} (1980) 203}.

\end{thebibliography}
\end{document}